\documentclass[11pt,a4paper]{article}
\usepackage[round]{natbib} 
\usepackage{amsmath} 
\numberwithin{equation}{section} 
\usepackage{graphicx} 
\usepackage{here} 
\usepackage{enumitem} 
\usepackage[font=footnotesize]{caption} 

\usepackage{Sweave}
\begin{document}
\title{Nested Archimedean copulas: \\ a new class of nonparametric tree structure estimators}
\author{
    Nathan Uyttendaele\footnote{Universit\'e catholique de Louvain, Institut de Statistique, Biostatistique et Sciences Actuarielles, Voie du Roman Pays 20, B-1348 Louvain-la-Neuve, Belgium.} \\
    \small nathan.uyttendaele@uclouvain.be
}
\date{\today}
\maketitle

\begin{abstract}
Any nested Archimedean copula is defined starting from a rooted phylogenetic tree, for which a new class of nonparametric estimators is presented. An estimator from this new class relies on a two-step procedure where first a binary tree is built and second is collapsed if necessary to give an estimate of the target tree structure. Several examples of estimators from this class are given and the performance of each of these estimators, as well as of the only known comparable estimator, is assessed by means of a simulation study involving target structures in various dimensions, showing that the new estimators, besides being faster, usually offer better performance as well. Further, among the given examples of estimators from the new class, one of the best performing one is applied on three datasets: 482 students and their results to various examens, 26 European countries in 1979 and the percentage of workers employed in different economic activities, and 104 countries in 2002 for which various health-related variables are available. The resulting estimated trees offer valuable insights on the analyzed data. The future of nested Archimedean copulas in general is also discussed.
\end{abstract}

\newpage
\section{Introduction \label{introduction}}
Archimedean copulas (ACs) have become a standard tool for modelling or simulating bivariate data. They are among the more popular classes of copulas, if not the most popular one. Their success story however falls short as soon as they are used to model or simulate higher-dimensional datasets. Indeed, because of their functional form, Archimedean copulas usually fail to properly take into account the dependencies between more than two random variables, as briefly discussed in the next section.

Nested Archimedean copulas (NACs), also called hierarchical Archimedean copulas (HACs), introduced by \citet[pp.~87--89]{tJOE97a}, are a natural generalization of Archimedean copulas. The key feature of an Archimedean copula is its generator, which can be loosely defined as a function of a single argument. Nested Archimedean copulas are made up of two parts: a rooted tree structure and a collection of generators. They offer more flexibility for modelling dependencies in a high-dimensional setting while still reducing to Archimedean copulas in simpler cases.

\cite{segers2014nonparametric} presented a nonparametric NAC tree structure estimator and assessed its performance by means of a simulation study involving target structures spanned on up to seven random variables. The term nonparametric refers to the ability of their estimator to be used without making a single assumption about the target NAC prior to the estimation of its structure, in contrast to the tree structure estimator published by \cite{OOW}.

After a short introduction to Archimedean copulas and nested Archimedean copulas in Section \ref{ACNAC}, including a summary of the key points of \cite{segers2014nonparametric}, a new class of nonparametric tree structure estimators is presented in Section \ref{new_est}. Several examples of estimators from this class are given; they all rely on a two-step procedure where first, a binary tree is built and second, parts of the binary tree is collapsed according to a given criterion. Both steps are required to be carried out without making a single assumption about the target NAC prior to the estimation of its structure or, at worst, by making one weak assumption, allowing to describe estimators from this new class as nonparametric, too.

The performance of several estimators given as examples in Section \ref{new_est} is then assessed by means of a simulation study involving target structures spanned on up to forty random variables, see Section \ref{performances}. A subset of these target structures is also used to assess the performance of the nonparametric estimator from \cite{segers2014nonparametric}. It is ultimately concluded that, while the new nonparametric estimators do not form a homogeneous group regarding performance, they usually perform better than the nonparametric estimator developed by \cite{segers2014nonparametric} and are moreover faster.

In Section \ref{application}, one of the new estimators is applied on three datasets: 482 students and their results to various examens, 26 European countries in 1979 and the percentage of workers employed in different economic activities, and 104 countries in 2002 for which various health-related variables are available. Although the generators of the target NAC are not estimated, the estimated structure itself, with an estimated summary measure of the dependence at each internal node of the structure, allows for valuable insights on the analyzed data.

Finally, in Section \ref{discussion}, the problem of estimating the generators of a NAC is outlined, as well as some remaining challenges still preventing NACs to reach their full potential in the same way Archimedean copulas did.

\section{Nested Archimedean copulas \label{ACNAC}}
Let $(X_1, \dots, X_d)$ be a vector of continuous random variables. The copula of this vector is defined as
\begin{equation} \nonumber
C(u_1, \dots, u_d)=P(U_1\leq u_1, \dots, U_d\leq u_d),
\end{equation}
where $(U_1, \dots, U_d)=(F_{X_1}(X_1), \dots, F_{X_d}(X_d))$, and where $F_{X_1}, \dots, F_{X_d}$ are the marginal cumulative distribution functions (CDFs) of $X_1, \dots, X_d$, respectively.

Archimedean copulas (ACs) can always be written in closed form as
\begin{equation} \nonumber
C(u_1, \dots, u_d)=\psi(\psi^{-1}(u_1) +\dots+\psi^{-1}(u_d)),
\end{equation}
where $\psi$ is called the generator and $\psi^{-1}$ is its generalized inverse, with $\psi : [0, \infty) \rightarrow [0, 1]$, a convex, decreasing function such that $\psi(0)=1$ and $\psi(\infty)=0$. In order for $C$ to be a $d$-dimensional copula, the generator is required to be $d$-monotone on $[0, \infty)$, see \cite*{2009arXiv0908.3750M} for more details.

Estimation of an AC is usually performed either by assuming $\psi$ belongs to a parametric family (for a list of popular families, see for instance \citealp{Hofert:Maechler:2010:JSSOBK:v39i09}) or by not assuming anything about $\psi$, that is, the whole $\psi$ function has to be estimated, see \cite{genest2011inference}.

In the case of Archimedean copulas, $C(u_1, \dots, u_d)$ is a symmetric function in its arguments and this is why Archimedean copulas are sometimes called \emph{exchangeable}. The result of this exchangeability property is easily seen by plotting a cloud of points generated from a bivariate Archimedean copula: the $y=x$ axis is a clear axis of reflection symmetry for the underlying distribution. For a cloud of points generated from a trivariate Archimedean copula, even more complex symmetries for the underlying trivariate distribution can be observed.

Another consequence of this exchangeability property is that, given a $d$-variate Archimedean copula and $m \in \{2, \ldots, d-1\}$, any two $m$-variate margins from that Archimedean copula describe the same $m$-variate distribution. For instance, with $m=3$ and assuming the joint distribution of $(U_1, \dots, U_{10})$ is an Archimedean copula, the joint distribution of $(U_6, U_5, U_3)$ is equal to the joint distribution of $(U_3, U_{10}, U_2)$ or $(U_1, U_4, U_8)$.

It is clear that, for modelling purposes, this exchangeability property becomes an increasingly strong assumption as the dimension $d$ grows.

Nested Archimedean copulas allow to relax this exchangeability property. They are obtained by plugging in Archimedean copulas into each other \citep[pp.~87--89]{tJOE97a}. The following example shows how a bivariate Archimedean copula $C_{23}$ can be plugged into a bivariate Archimedean copula $C_{123}$:
\begin{multline} \label{B_0}
C_{123}(u_1, \boldsymbol{C_{23}(u_2, u_3)})= \\ \psi_{123}\big(\psi_{123}^{-1}(u_1) +\psi_{123}^{-1}(\boldsymbol{\psi_{23}(\psi_{23}^{-1}(u_2) +\psi_{23}^{-1}(u_3))})\big)
\end{multline}

The above trivariate copula, a nested Archimedean copula on $(U_1, U_2, U_3)$, is still such that the $y=x$ axis remains an axis of reflection symmetry for all bivariate margins. However, while even more complex symmetries are observed on the trivariate level for an AC, part of these symmetries are lost on the trivariate level of the NAC described by (\ref{B_0}). Moreover, while all bivariate margins are the same in an AC, the distribution of ($U_2$, $U_3$) is not the same as the distribution of ($U_1$, $U_2$) or ($U_1$, $U_3$) in the NAC described by (\ref{B_0}). All these remarks hold provided the generators $\psi_{123}$ and $\psi_{{23}}$ are different, for otherwise (\ref{B_0}) can be simplified back to a trivariate AC. In general, the more the generators of successive nodes from a NAC are different, the more that NAC will be said to be \emph{resolved}, a word mainly used in Section \ref{performances}. Poorly resolved NACs are almost ACs.

\begin{figure}[H]
\centering
\begin{tabular}{cc}

\includegraphics[width=0.2\textwidth]{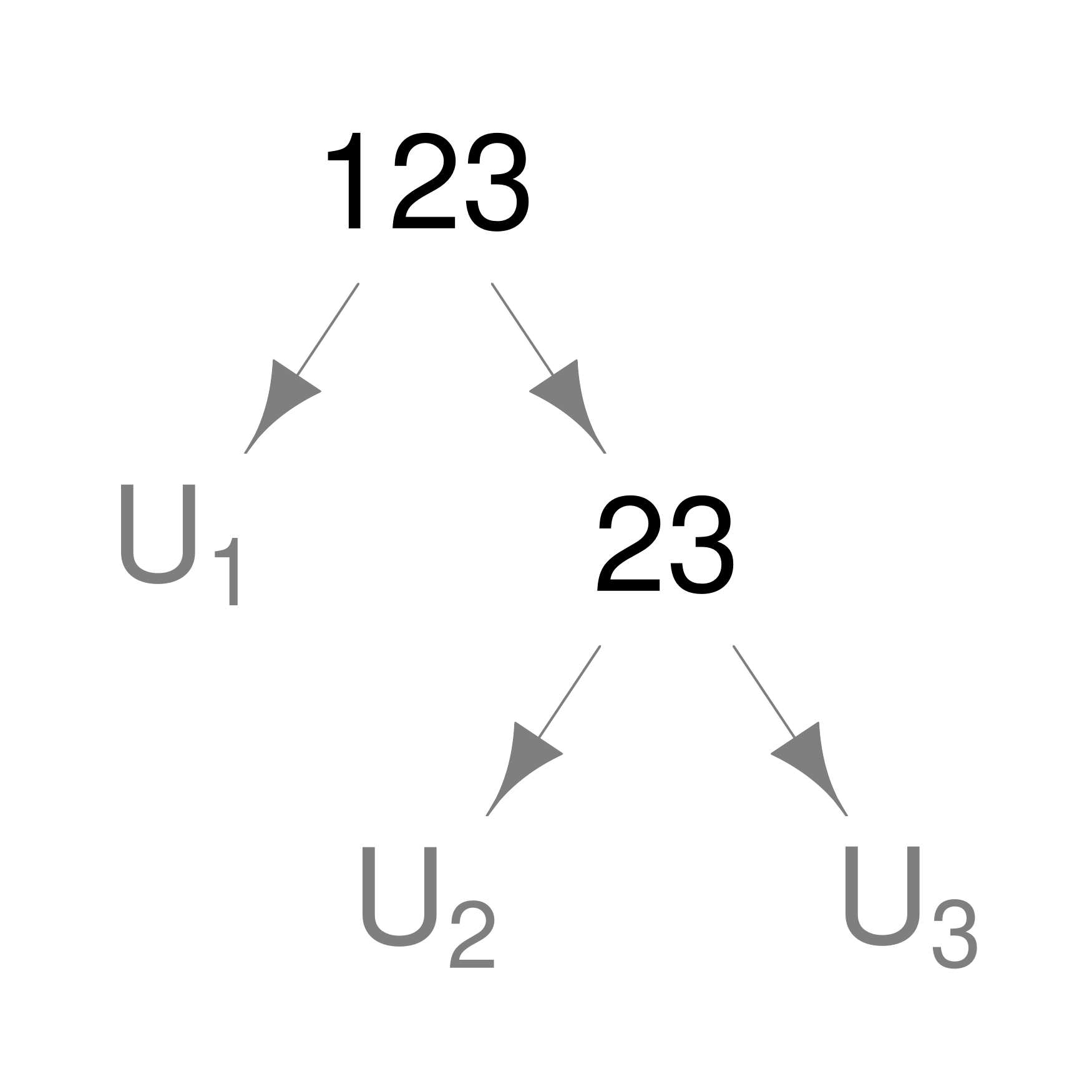}
&
\includegraphics[width=0.2\textwidth]{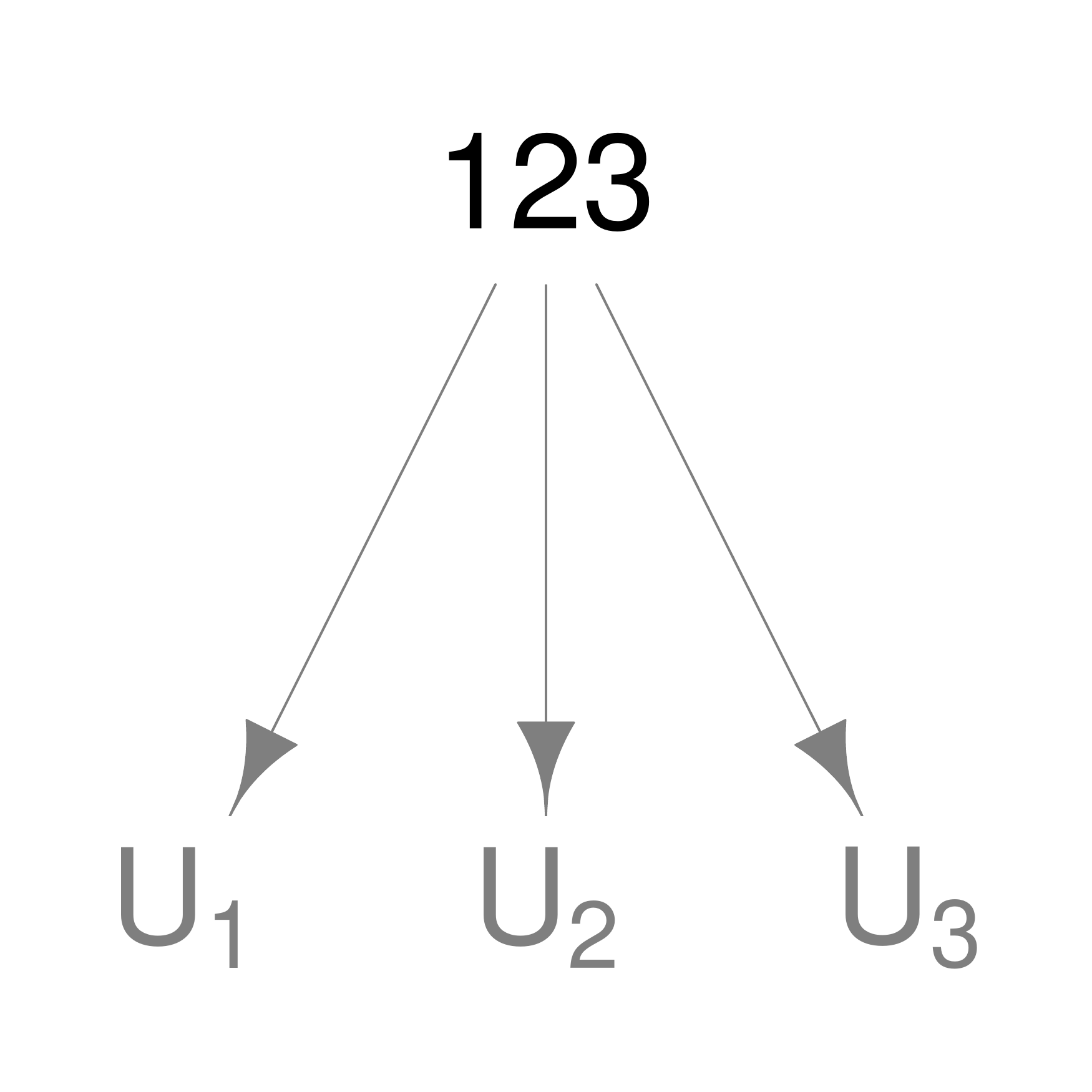}

\end{tabular}
\caption{Left: the tree structure implied by (\ref{B_0}). The dependence between $U_2$ and $U_3$ is described by $\psi_{23}$ while the dependence between $U_1$ and $U_2$ or $U_1$ and $U_3$ is described by $\psi_{123}$. Right: the structure of a trivariate AC. The dependence between any two random variables is here described by $\psi_{123}$. The arrows in both structures are called edges. Note that the labels of the internal nodes in both structures are actually irrelevant. \label{tri_struct}}
\end{figure}

In a NAC, the way Archimedean copulas are nested corresponds to a rooted tree. This tree is such that any internal node, an internal node being any node different from a leaf, must have at least two children and every node but the root must have one and only parent. Nested Archimedean copulas, such as the one in (\ref{B_0}) and for which the tree can be seen on the left panel of Figure~\ref{tri_struct}, are defined through that rooted tree structure and through a collection of generators, one for each internal node in the tree. Each generator fully describes the dependence between the random variables interacting through the related node. Archimedean copulas can be seen as a special case of NACs: they exhibit a trivial structure such as the one on the right-hand panel of Figure \ref{tri_struct} and have only one generator, the one related to the only internal node, the root. A trivial tree structure of dimension $d$ is sometimes called a $d$-fan (\citealp{ng1996}).

\cite{segers2014nonparametric} further explore NAC trees. One of their key findings is the following: the tree structure of a nested Archimedean copula, spanned on a vector of continuous random variables $(U_1, \ldots, U_d)$, can be uniquely retrieved provided the tree structure of the (trivariate) nested Archimedean copula associated with any subset of three distinct random variables of $(U_1, \ldots, U_d)$ is known.

Simply stated, any tree structure $\lambda$ can be broken down into a unique set, denoted by $^{3}(\lambda)$, consisting of $\bigl(\begin{smallmatrix}
d \\ 3
\end{smallmatrix} \bigr)$ trivariate tree structures, one trivariate structure for each combination of the elements of $(U_1, \ldots, U_d)$, taken three at the time without repetition. Moreover, a given set $^{3}(\lambda)$ can in turn be used to retrieve the tree structure $\lambda$ from which it originated.

To estimate a tree structure $\lambda$, \cite{segers2014nonparametric} therefore suggest to estimate, one at the time, each element of $^{3}(\lambda)$, thus effectively getting $\widehat{^{3}(\lambda)}$ which can then be used to build $\hat{\lambda}$.

The ability to estimate the tree structure spanned on three random variables $(U_i, U_j, U_k)$ based on $n$ observations from $(X_i, X_j, X_k)$ is a critical requirement for the estimation of $^{3}(\lambda)$. As outlined by \cite{segers2014nonparametric}, there are in fact only four possible structures in the trivariate case: a trivial structure, such as the one on the right-hand side of Figure \ref{tri_struct}, or a structure where one variable is left apart and the two others are put together, as seen on the left-hand side of Figure \ref{tri_struct}, where $U_2$ and $U_3$ are put together and $U_1$ is located closer to the root.

If the trivariate target structure is assumed not to be the trivial structure, then picking one of the three remaining structures as estimate of the trivariate target structure is not a complicated problem: just estimate the Kendall distribution for each of the tree pairs $(X_i, X_j)$, $(X_i, X_k)$ and $(X_j, X_k)$, and find out which are the two estimated Kendall distributions that are the closest according to some distance. If, for instance, the estimated Kendall distributions of $(X_i, X_j)$ and $(X_i, X_k)$ are the closest, then the trivariate target tree structure must be a structure where $U_i$ is left apart while $U_j$ and $U_k$ are together.

Please note that if for some reason the target structure $\lambda$ spanned on $(U_1, \ldots, U_n)$ is known to be a binary structure, then so is each element of $^{3}(\lambda)$. Therefore, in this particular case, each element of $^{3}(\lambda)$ can be estimated using only what is described in the previous paragraph.

Finding out if a trivariate target structure is actually the trivial structure or not is a much more difficult problem, for which \cite{segers2014nonparametric} developed a hypothesis test where they try to see if the average of the two closest estimated Kendall distributions is significantly different from the third estimated Kendall distribution. If it is not the case, then it is not possible to rule out that the three underlying Kendall distributions all coincide, and the trivariate target structure is estimated by a 3-fan. The distribution of their test statistic being unknown under the null, they rely on the bootstrap to get a p-value for the test. As the estimation of all elements from $^{3}(\lambda)$ will require this test to be performed $\bigl(\begin{smallmatrix}
d \\ 3
\end{smallmatrix} \bigr)$ times, getting $\hat{\lambda}$ using their approach can be computationally intensive, especially as the value of $d$ increases.

Some suggested papers for the readers eager to learn more about NACs are: \cite{doi:10.1080/00949650701255834}, \cite{hofert2013densities} or \cite{okhrinOstap2013properties}.

\section{A new class of nonparametric tree structure estimators \label{new_est}}
In this section, a new class of NAC tree structure estimators is presented, with a few examples. An estimator from this class always consists in two steps: a first step where a binary tree is built and a second step where it is collapsed if necessary.

{\bf Step one}. It is first assumed that the target structure spanned on $(U_1, \ldots, U_d)$ is a binary tree, that is, a structure where each internal node has two and only two children (this assumption will be later relaxed). Note the tree on the left-hand side of Figure \ref{tri_struct} is actually a binary tree, the smallest binary tree possible.

Based on an iid sample of size $n$ from $(X_1, ..., X_d)$ and knowing that the tree structure spanned on $(U_1,\ldots, U_d)$ is a binary tree, a distance for each couple $(X_i, X_j)$ with distinct $i, j \in \{1, \ldots, d\}$ is first estimated. The random variables are then clustered, one at the time, according to the estimated distances. Getting an estimated binary tree structure this way however makes sense only if the estimated distance for each couple $(X_i, X_j)$ is a measure of dependence (a large dependence being translated by a small distance) and if the target NAC is such that the dependence between any two random variables in the structure increases as the variables are able to interact through nodes that are farther away from the root (loosely said, the dependence between random variables increases as one goes down the structure).

It is unclear if this last assumption, in general, always holds for a NAC. However, should this assumption not be true in general, it is important to note it remains a weak assumption about the target NAC.

A measure of dependence between two random variables can be obtained, for instance, through
\begin{itemize}[noitemsep, nolistsep]
\item Kendall's $\tau$,
\item a distance between the (theoretical) Kendall distribution of two independent variables and the empirical Kendall distribution of the two variables under study,
\item or Hoeffding's $D$ statistic (see \citealp{hoeffding1948non}).
\end{itemize}

Notice these distances are all such that $\text{dist}(X_i, X_j)=\text{dist}(U_i, U_j)$, so that the binary tree estimated on $(X_1, ..., X_d)$ is actually the binary tree estimated on $(U_1,\ldots, U_d)$.

To cluster the random variables one at the time, usual clustering techniques such as single or average linkage can be used. In the rest of this paper, only average linkage will be considered.

To avoid the assumption of increasing dependence as one goes down the structure, step one can alternatively be carried out using a \emph{supertree method}. Supertree methods are designed to output a structure (called a supertree) that will represent as well as possible an input set of smaller trees, this input set including trees of various sizes, conflicting trees and also missing trees (that is, some information to build the representative structure is actually lacking). Supertree methods have been extensively studied in the field of phylogenetics. Some interesting references to get started are \cite{bininda2004evolution}, \cite{wilkinson2005shape} or \cite{swenson2012superfine}.

In this paper, two supertree methods implemented in the \textsf{R} package \textsf{phytools} (\citealp{phyliam}) have been used. They both take as input a set of unrooted trees and output an unrooted supertree. Hereafter is described how they work and also how to make sure the outputted supertree can be rooted in a meaningful way.

To start, and assuming all the input trees are unrooted (if they are not, they are first unrooted), all the topological information available across the input trees is gathered as a matrix, called the character matrix. This matrix contains as many rows as there are leaves in the supertree one wants to build and as many columns as there are internal edges in all the input trees. Given an internal edge (say, the first column of the matrix) the leaves from the related input tree can be seen as divided into two sets. Visualize the  internal edge under consideration as an horizontal line: some leaves are going to be on its left, some are going to be on its right. Leaves receive a label in the first column depending on which side of the internal edge under consideration they are. Leaves that are not part of the input tree related to the internal edge under consideration receive what is called an unknown state.

\begin{figure}[H]
\centering
\begin{minipage}[c]{0.275\textwidth}
\includegraphics[width=1\textwidth]{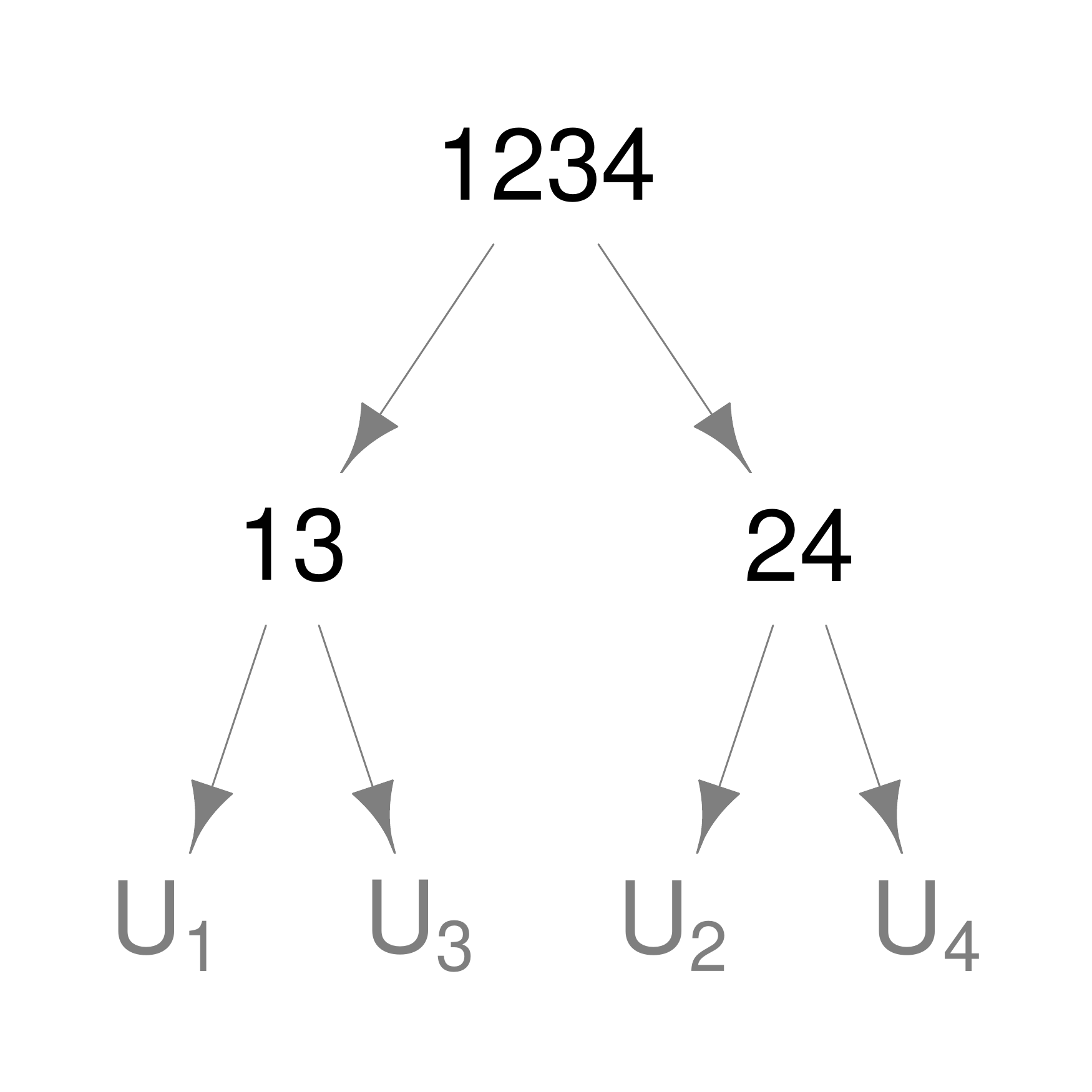}
\end{minipage}
\begin{minipage}[c]{0.275\textwidth}
\includegraphics[width=1\textwidth]{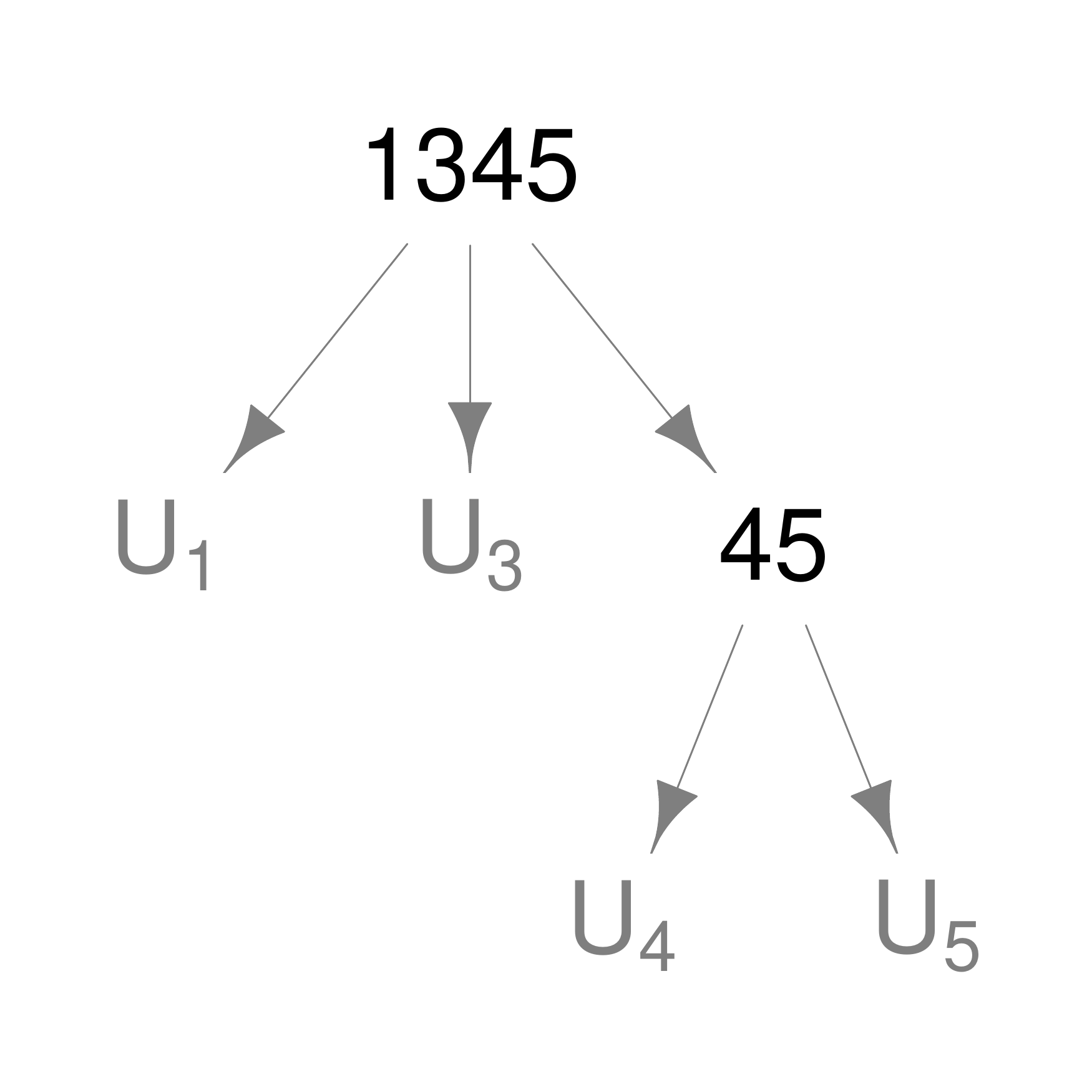}
\end{minipage}
\begin{minipage}[c]{0.15\textwidth}
\centering
\begin{tabular}{c|cc}
  $U_1$ & 0 & 0 \\
  $U_2$ & 1 & ? \\
  $U_3$ & 0 & 0 \\
  $U_4$ & 1 & 1 \\
  $U_5$ & ? & 1
\end{tabular}
\end{minipage}
\begin{minipage}[c]{0.25\textwidth}
\centering
\begin{tabular}{c|ccc}
  $U_1$ & 0 & 1 & 0 \\
  $U_2$ & 1 & 0 & ? \\
 $U_3$ & 0 & 1 & 0 \\
  $U_4$ & 1 & 0 & 1 \\
  $U_5$ & ? & ? & 1 \\
  $O$ & 0 & 0 & 0
\end{tabular}
\end{minipage}
\caption{Two rooted input trees and the related character matrix built on the unrooted version of these two trees first without and second with an outgroup added.\label{input_MRP_trees}}
\end{figure}

As an example, consider the two tree structures in Figure \ref{input_MRP_trees}. Suppose one wants to build a tree spanned on $(U_1, \ldots, U_5)$ based on these two trees. Both trees are first unrooted. The first rooted tree in Figure \ref{input_MRP_trees} loses one internal edge in the process and becomes a tree with structure of the form >$-$<, where the only remaining internal edge has nodes 13 and 24 at its tips. The second rooted tree in Figure \ref{input_MRP_trees}, when unrooted, also becomes a tree with structure of the form >$-$<, where the internal edge has nodes 1345 and 45 at its tips. The character matrix that gathers all the topological information available based on these two unrooted trees is the $5\times2$ matrix displayed in Figure \ref{input_MRP_trees}. Looking at the internal edge in the first unrooted tree, we see that $U_1$ and $U_3$ are on one side of that edge and $U_2$ and $U_4$ are on the other side. We assign the label (or state) 0 to $U_1$ and $U_3$, and the label 1 to $U_4$ and $U_2$. The leaf $U_5$ does not appear in this input unrooted tree and it is therefore not known to which side this leaf belongs. We therefore assign an unknown state to $U_5$. For the second unrooted tree, $U_1$ and $U_3$ are on one side of the internal edge and $U_4$ and $U_5$ are on the other side. We do not know to which side $U_2$ belongs.

Once the character matrix has been build, the next step is to find an unrooted supertree that will be in agreement as much as possible with the topological information available in the matrix. As loss function, phylogeneticians use what is called the parsimony score, which can be easily calculated for a given supertree using the algorithm developed by \cite{fitch1971}.

To find the supertree with the minimum parsimony score, the strategy is to pick a starting supertree and then to apply topological rearrangements to that supertree in a recursive fashion. Rearrangements leading to a supertree with a lower parsimony score are kept, changes such that the resulting supertree has a higher parsimony score are not kept. The final supertree is one such that no further rearrangements of the supertree allows to lower the parsimony score.

The two supertree methods tested in this paper differ only in the way the starting supertree is defined and in the way the starting tree is recursively modified. The first of these methods, later denoted by \textsf{NJNNI}, uses as starting supertree a tree built based on the neighbor joining (NJ) clustering method from \cite{saitou1987neighbor}. The changes applied recursively on the tree are NNI rearrangements, see \citet[pp. 39]{felsenstein2004inferring} for a detailed description of such rearrangement.

For the second method, later denoted by \textsf{RNix}, the starting tree is chosen at random and is then modified according to \cite{nixon1999parsimony}.

Both these methods output an unrooted supertree and, by unrooting input rooted trees, destroy the topological information that could be used to root the outputted supertree. To avoid this, a leaf, called the outgroup, is attached to the root of each input rooted tree prior to their unrooting. As an example, the character matrix with such outgroup for the two input trees in Figure \ref{input_MRP_trees} is displayed on the most right part of the same figure. Notice this character matrix, with outgroup, has now as many columns as there are edges in the original, rooted, input trees. The supertree based on this $6\times3$ character matrix will be spanned on $(U_1, \ldots, U_5)$ and the outgroup. The final step is to use the outgroup to root the supertree before removing the outgroup.

In order to estimate a target binary NAC tree structure using one of the two supertree methods described above, the suggestion is to use, as input set of trees, the set of binary trivariate trees one gets by estimating the binary tree spanned on each vector of three random variables $(U_i, U_j, U_k)$ with distinct $i, j, k \in \{1, \ldots, d\}$, refer to Section \ref{ACNAC} for more details.

{\bf Step two}. In any case, whether a supertree method is used or not to estimate the target structure, the assumption that the target tree structure is a binary structure is made. However, a target NAC structure is not necessarily a binary structure. To allow for a more general estimation, the suggestion is to collapse, if necessary, one or several parts of the estimated binary structure.

\begin{figure}[H]
\centering
\begin{tabular}{cc}

\includegraphics[width=0.275\textwidth]{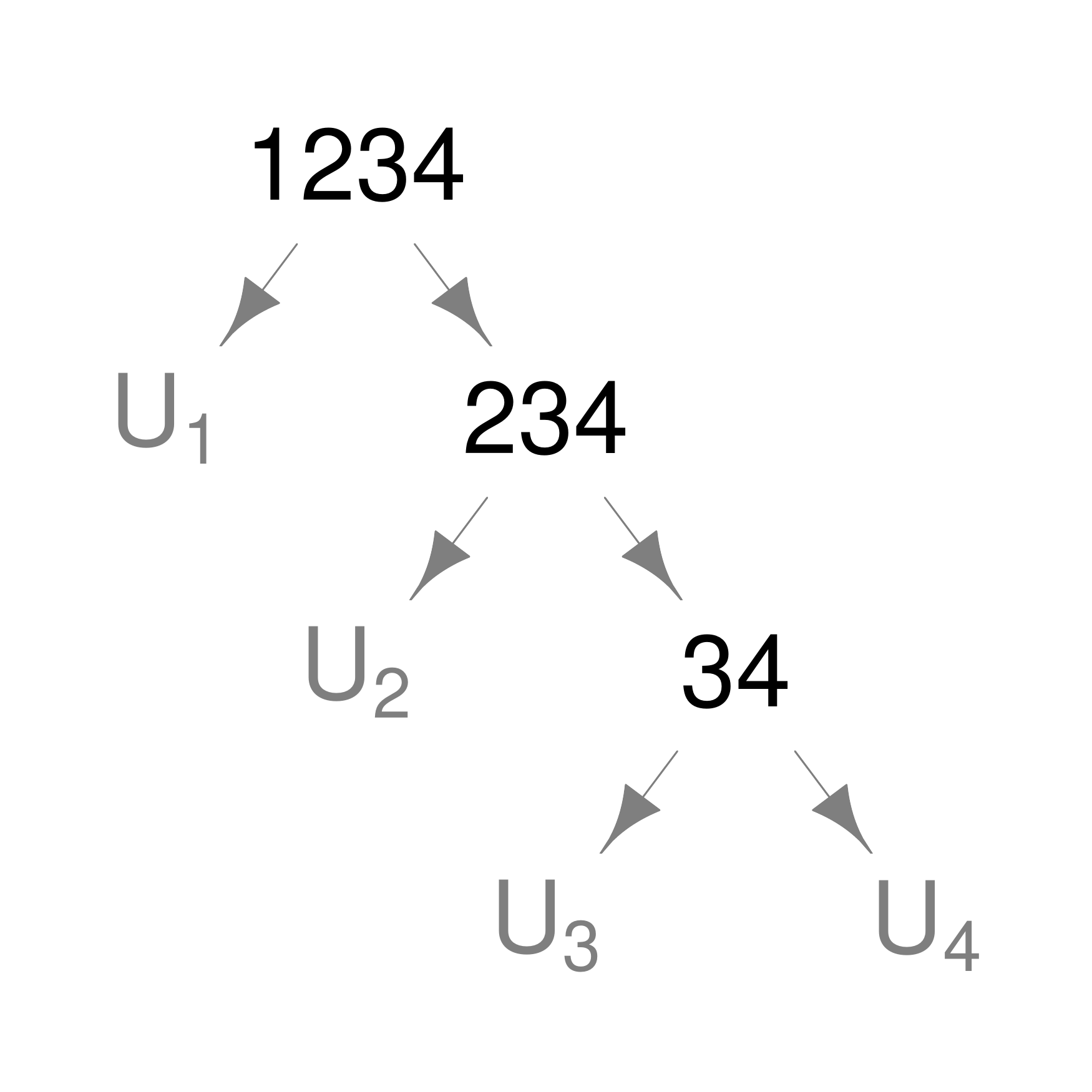}
&
\includegraphics[width=0.275\textwidth]{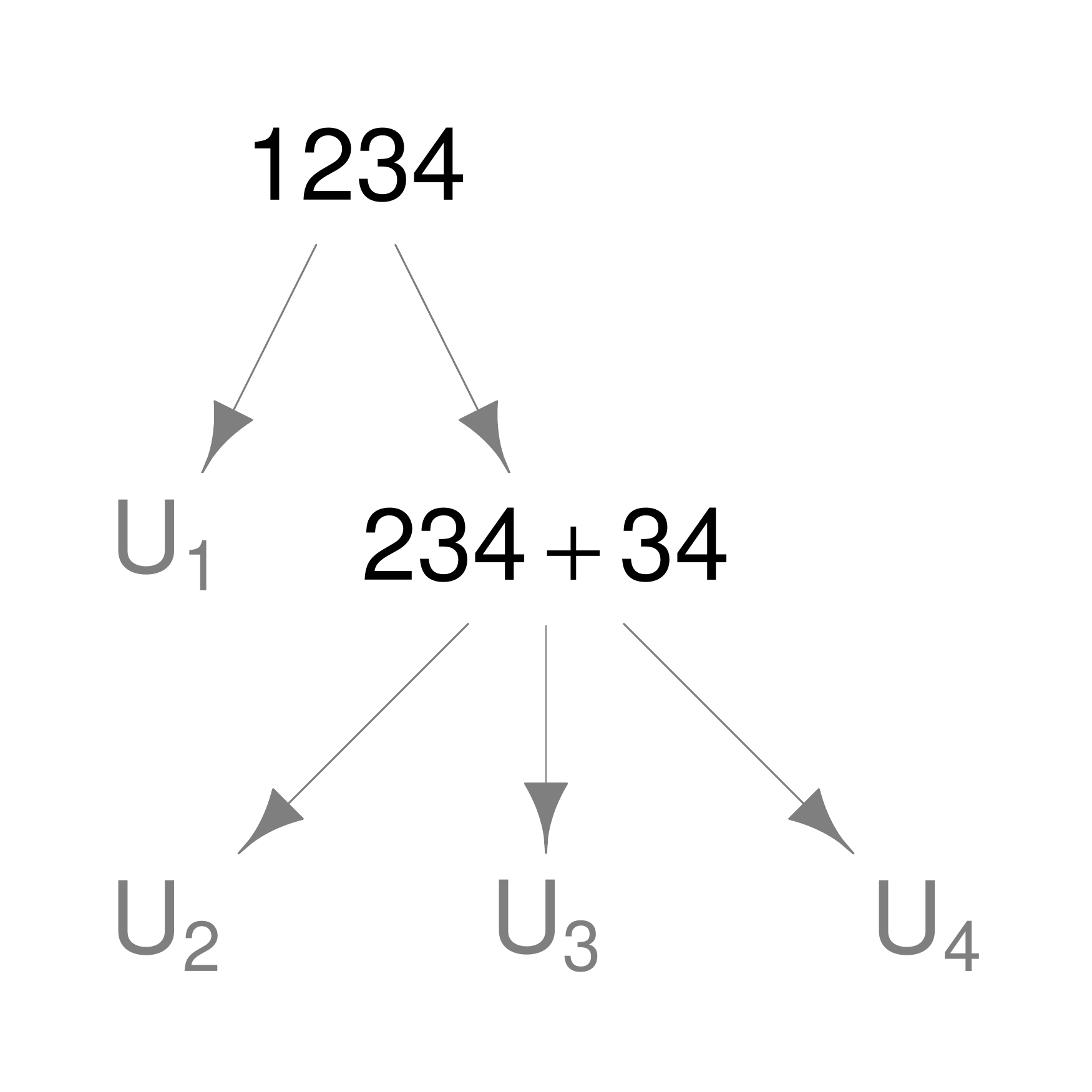}

\end{tabular}
\caption{Left: a binary structure spanned on four random variables. Right: nodes $234$ and $34$ have been collapsed into one node. \label{collapse}}
\end{figure}

Take for instance the estimated binary tree displayed on the left-hand side of Figure~\ref{collapse}. Since each internal node is related to a generator describing the dependence between the random variables interacting through that node, it makes sense to check if the generator at a given internal node is not too similar to the generator at the previous or next internal node in the structure. If it turns out to be the case, then the nodes in question should be collapsed into a single node to avoid overfitting. The right-hand side of Figure \ref{collapse} shows the collapsing of nodes $234$ and $34$ into a new node. Notice the resulting tree is not a binary tree anymore.

To compare two successive generators, one needs first to estimate these generators, as they are unknown. To make this problem easier, the suggestion is to summarize each generator as a scalar measure reflecting the dependence between the random variables interacting through the related node and to collapse two successive nodes if the absolute difference between their respective estimated scalar measure is lower than a chosen threshold.

For instance, looking back at Figure \ref{collapse}, one can estimate a summary of the dependence between the random variables interacting through node $234$ by averaging the estimated Kendall's $\tau$s within the random pairs $(U_2, U_3)$ and $(U_2, U_4)$. The estimated summary of the generator related to node $34$ is equal to the estimated Kendall's $\tau$ between the random variables $(U_3, U_4)$. The two nodes are collapsed if the inequality

\begin{equation} \label{critical}
\left|\left(\frac{\hat{\tau}_{23}+\hat{\tau}_{24}}{2}\right)-\hat{\tau}_{34}\right|<\tau_c
\end{equation}
holds, where $\tau_c$ is the critical threshold for collapsing.

Instead of using Kendall's $\tau$ in (\ref{critical}), other measures of dependence can be used, for instance Spearman's $\rho$, Hoeffding's $D$ statistic, etc.

Another suggestion to decide if two successive nodes should be collapsed or not is to break down both structures before and after collapsing of two given nodes into their respective set of trivariate pieces (refer to Section~\ref{ACNAC}). Since the structures before and after collapsing are different, some of the trivariate pieces will be different as well. More precisely, some of the trivariate pieces before collapsing will be binary trees, while after collapsing they become 3-fans.
Looking back at Figure~\ref{collapse}, one can observe that, in the left structure, the tree spanned on $(U_2, U_3, U_4)$ is not a $3$-fan, while in the right structure it is. The decision to collapse or not is made by performing the hypothesis test developed by \cite{segers2014nonparametric}, since this test precisely aims at deciding whether or not the tree structure spanned on three random variables $(U_i, U_j, U_k)$ is a trivial tree structure or not. If the p-value of the test is lower than or equal to a threshold $\alpha$, the nodes $234$ and $34$ should not be collapsed into one.

Things are harder if two or more vectors of the form $(U_i, U_j, U_k)$ do not have the same trivariate structure before and after collapsing. In such situations, the hypothesis test developed by \cite{segers2014nonparametric} has to be applied as many times as there are different trivariate pieces and it can happen that some trivariate pieces of the structure before collapsing are supported by the data (the trivial pieces in the structure after collapsing have been rejected) but some are not (the trivial pieces in the structure after collapsing have not been rejected). For such cases, a rule of thumb such as the one hereafter can be used (do not collapse if the inequality holds):

\begin{equation} \label{critical2}
\text{average p-value}\leq\alpha
\end{equation}


Estimators from the new class described in this section all have in common a first step where a binary tree is built and a second step where parts of the binary tree from step 1 are collapsed, if necessary, according to some criterion such as (\ref{critical2}) or (\ref{critical}). As emphasized in Section \ref{introduction}, both steps are further required to be carried out without making a single assumption about the target NAC prior to the estimation of its structure or, at worst, by making one weak assumption, allowing to describe estimators from this new class as nonparametric. As there are many ways to build a binary tree and even more ways to collapse a binary tree, this class encompasses a very large number of estimators. While it would be interesting to perform a large scale study of these estimators, the performance of only a handful of them will be studied in the next section.

Optimized \textsf{R} codes for the tree structure estimators described in this section are available by simple mail request to the author of this paper. These \textsf{R} codes will eventually be bundled into an \textsf{R} package.

\section{Performance study \label{performances}}
When the sample size is $n$, the methodology used in this section to estimate the performance of a NAC tree structure estimator with respect to a given NAC, defined through a target tree structure $\lambda$ and an arbitrary set $\Psi$ of generators, is described through the following steps:

\begin{itemize}[noitemsep, nolistsep]
\item generate $N=100$ samples of size $n$ from ($\lambda$, $\Psi$);
\item apply the NAC tree structure estimator on each of these $N$ samples while considering the univariate margins unknown and get $N$ estimates of $\lambda$;
\item calculate a distance between each estimate of $\lambda$ and $\lambda$ itself;
\item get the average of the $N$ resulting distances or some other descriptive measure of these distances, such as:
\begin{equation}
 (\text{average of the distances})^2 + \text{variance of the distances;} \label{desc}
\end{equation}

\item the lower the average of the distances or the lower (\ref{desc}), the better the performance of the estimator for the NAC defined through ($\lambda$, $\Psi$) and when the sample size is $n$.
\end{itemize}

Two distances between a given estimate of $\lambda$ and $\lambda$ itself are considered. The first one is a \textsf{01}-distance: if the estimate of $\lambda$ is actually equal to $\lambda$, then the distance is 0. Otherwise, the distance is 1. The second distance, called the \textsf{tri}-distance, is based on the comparison of the trivariate pieces of the estimate of $\lambda$ and the trivariate pieces of $\lambda$ itself. If both trees are equal, all the trivariate pieces will be equal as well, and the distance is 0. If among the $\bigl(\begin{smallmatrix}
d \\ 3
\end{smallmatrix} \bigr)$ trivariate pieces from the estimate of $\lambda$ and the $\bigl(\begin{smallmatrix}
d \\ 3
\end{smallmatrix} \bigr)$ trivariate pieces from $\lambda$ a total of $k$ pieces differ, then the distance is $k$. The maximum possible tri-distance is therefore $\bigl(\begin{smallmatrix}
d \\ 3
\end{smallmatrix} \bigr)$. Unlike the 01-distance, this last distance allows to assess how far from the target structure a misspecified structure produced by a given estimator is.

Since the estimator from \cite{segers2014nonparametric} is based on a hypothesis test, it is required to choose a threshold $\alpha$ prior to the estimation of a target tree structure.

Regarding the estimators from the new class described in Section \ref{new_est}, the choice of a threshold to decide if any two successive nodes in the estimated binary tree structure should be collapsed is also required prior to the estimation of a target tree structure, as seen in (\ref{critical2}) or (\ref{critical}).

Comparison of the performance of different estimators is therefore a challenge, as the performance of a given estimator depends on the chosen threshold for that estimator. Given a sample size $n$, a target NAC ($\lambda$, $\Psi$) and a tree structure estimator, the suggestion is to use a threshold ensuring that $P(\hat{\lambda}_n=\lambda)$ is maximized. This particular threshold will be called the optimal threshold. By making use of the related optimal threshold for each estimator, one only takes into account the best performance of each estimator, which should allow for ``fair'' comparisons between estimators.

In the particular case $\lambda$ is a binary structure,

\begin{itemize}[noitemsep, nolistsep]
\item the estimator from \cite{segers2014nonparametric}, which is based on $\bigl(\begin{smallmatrix}
d \\ 3
\end{smallmatrix} \bigr)$ hypothesis tests, should always reject the nulls. If one null is not rejected, it means the final estimated structure contains at least a 3-fan, and therefore the estimate of $\lambda$ cannot be equal to $\lambda$ itself. Thus the threshold $\alpha$ should be set to 100\% or more so that all nulls are always rejected.
\item Regarding the estimators from the new class, in case $\lambda$ is a binary structure, $P(\hat{\lambda}_n=\lambda)$ is maximized if the collapsing step is skipped. This can be achieved by setting $\alpha$ to 100\% or more in (\ref{critical2}), and $\tau_c$ to 0 or less in (\ref{critical}).
\end{itemize}

Although unknown when $\lambda$ is not a binary structure, the optimal threshold for an estimator can be estimated. Indeed, given $N=100$ samples of size $n$, a target NAC ($\lambda$, $\Psi$) and a tree structure estimator, the estimated optimal threshold is the value such that the average of the $N=100$ 01-distances, one distance between each estimate of $\lambda$ and $\lambda$ itself, is minimal.

In this section, estimators from the new class are explicitly named according to the two steps on which they rely. For the first step, \textsf{RNix} and \textsf{NJNNI} refer to the two supertree methods described in Section \ref{new_est} while \textsf{kind}, \textsf{hD} and \textsf{kt} refer to distance matrices containing measures of deviation from the independent bivariate Kendall distribution, Hoeffding's $D$ statistics or Kendall's $\tau$s, respectively. For the second step, \textsf{kb} and \textsf{kagg} refer to (\ref{critical2}) and (\ref{critical}), respectively. The tree structure estimator developed by \cite{segers2014nonparametric} will be referred to as \textsf{S\&U}.

\begin{figure}[H]
\centering
\begin{minipage}[c]{0.18\textwidth}
\includegraphics[width=1\textwidth]{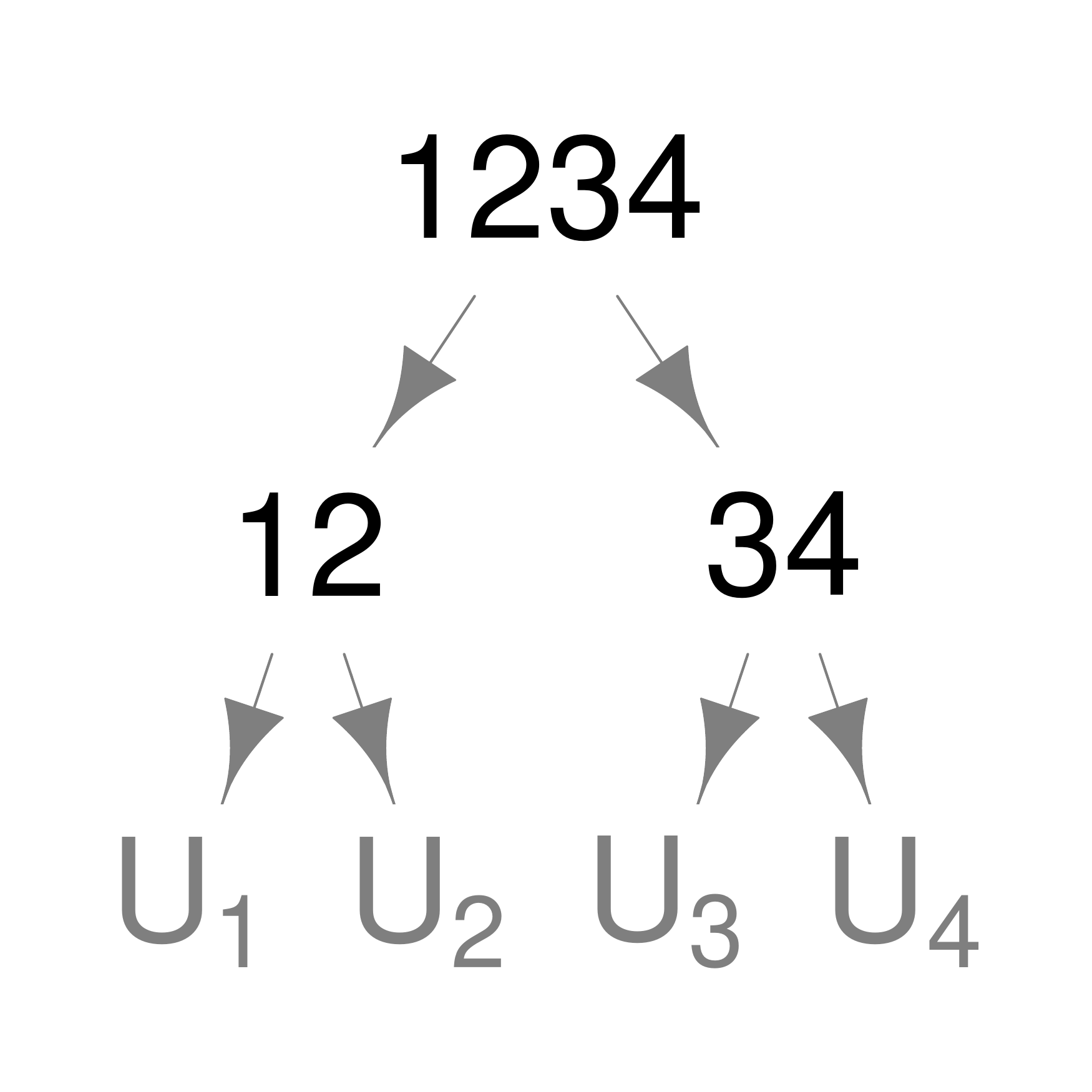}
\end{minipage}
\begin{minipage}[c]{0.18\textwidth}
\includegraphics[width=1\textwidth]{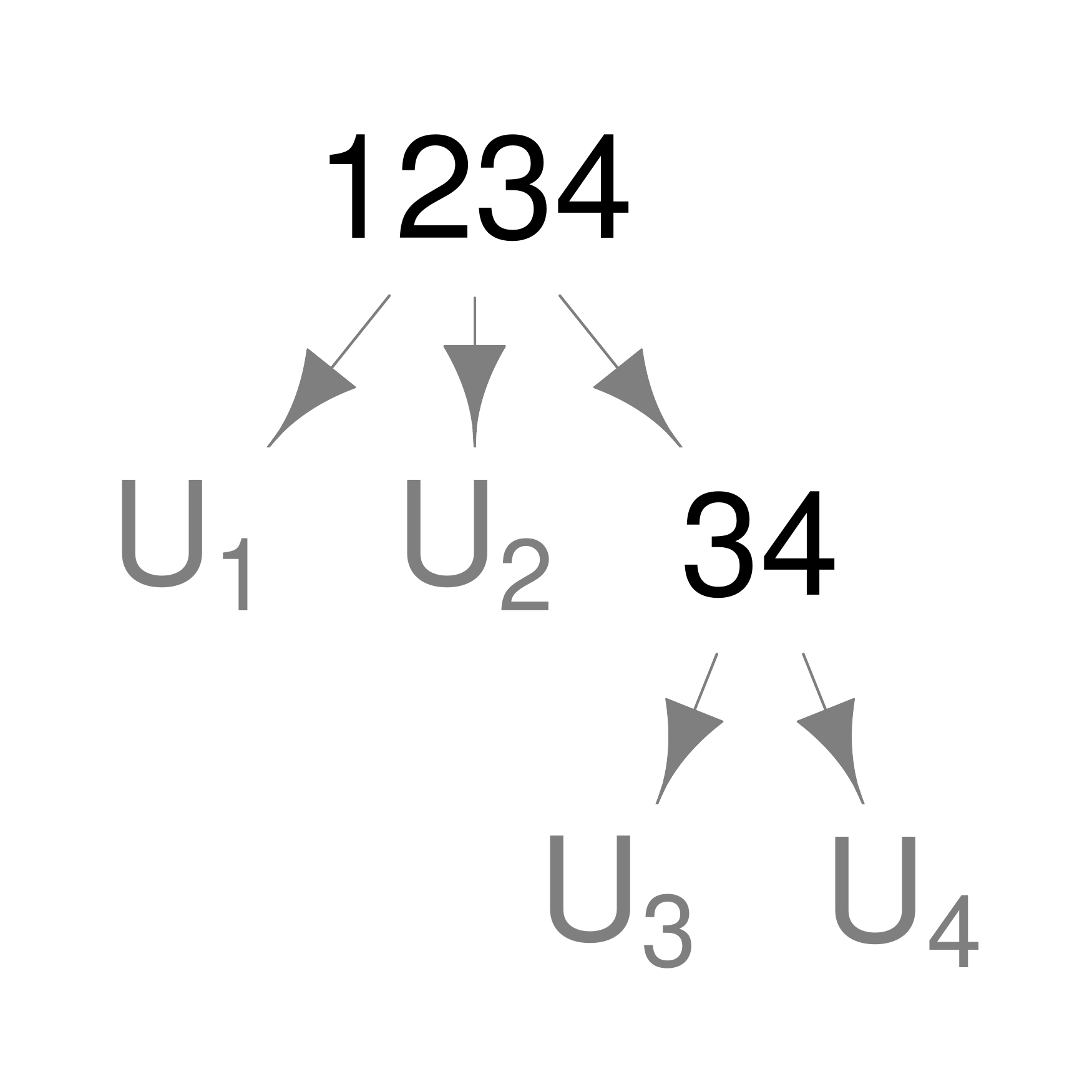}
\end{minipage}
\begin{minipage}[c]{0.26\textwidth}
\includegraphics[width=1\textwidth]{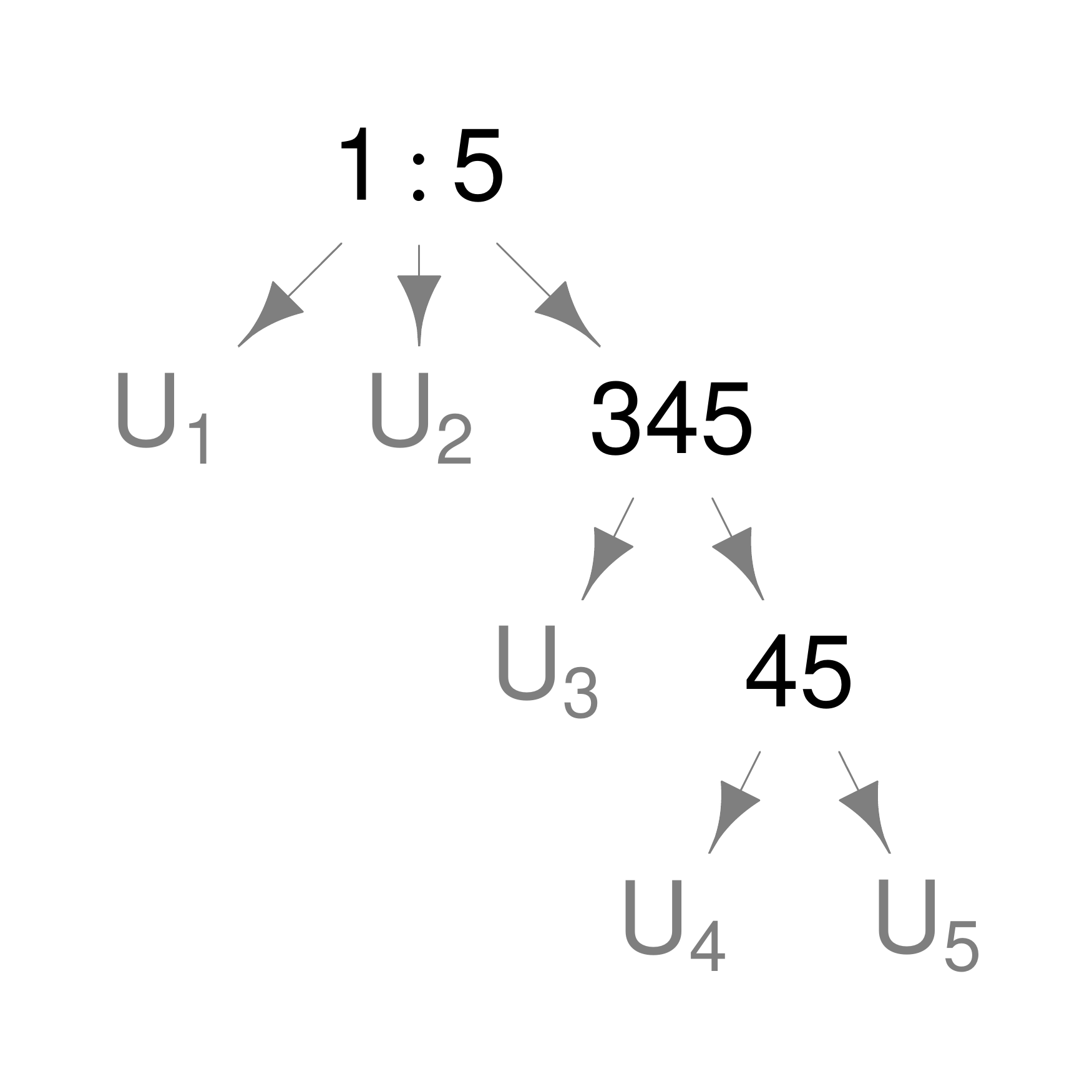}
\end{minipage}
\begin{minipage}[c]{0.35\textwidth}
\includegraphics[width=1\textwidth]{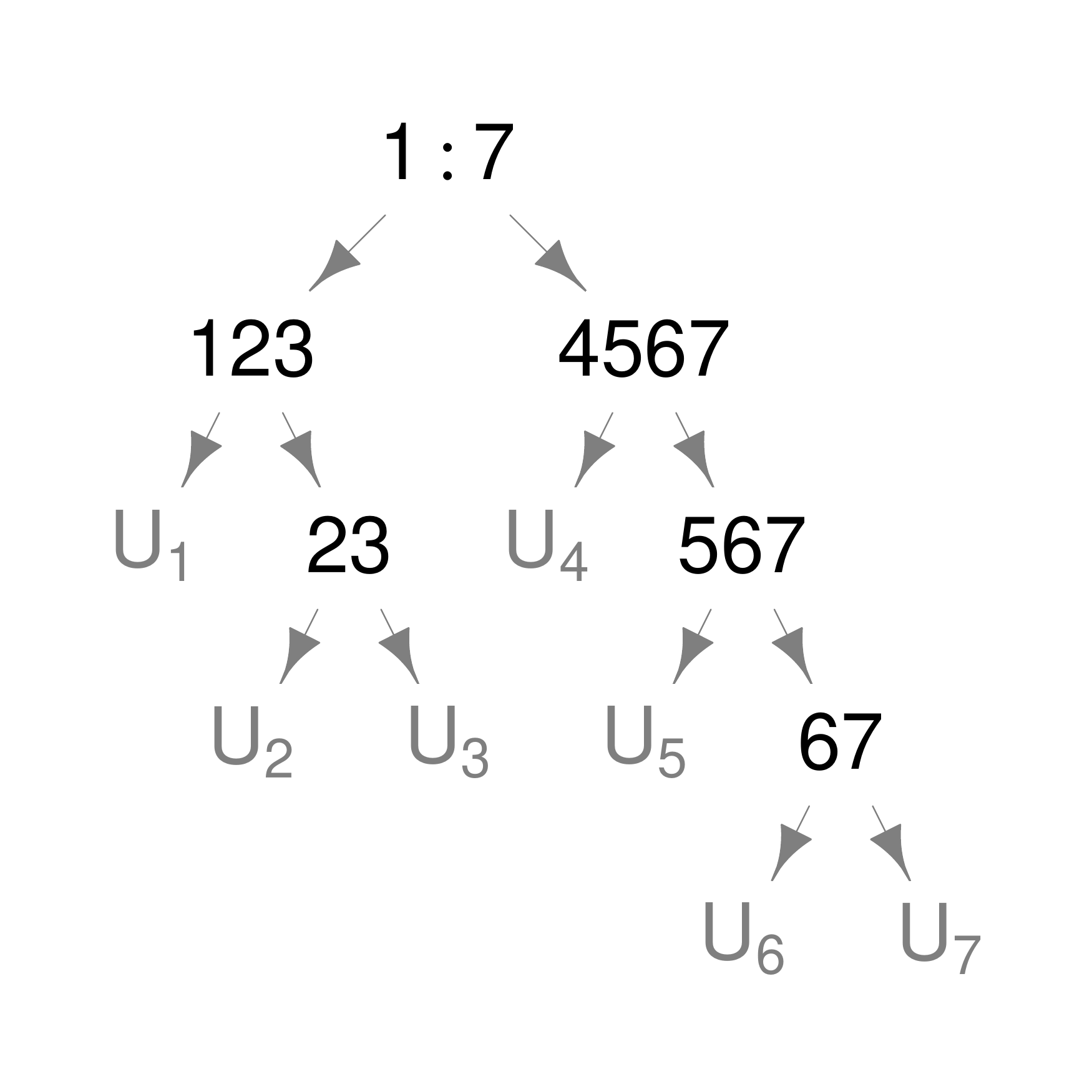}
\end{minipage}
\caption{two fourvariate structures, a fivevariate structure and a sevenvariate structure.\label{small_struct}}
\end{figure}

\begin{figure}[H]
\centering
\begin{minipage}[c]{0.4\textwidth}
\includegraphics[width=1\textwidth]{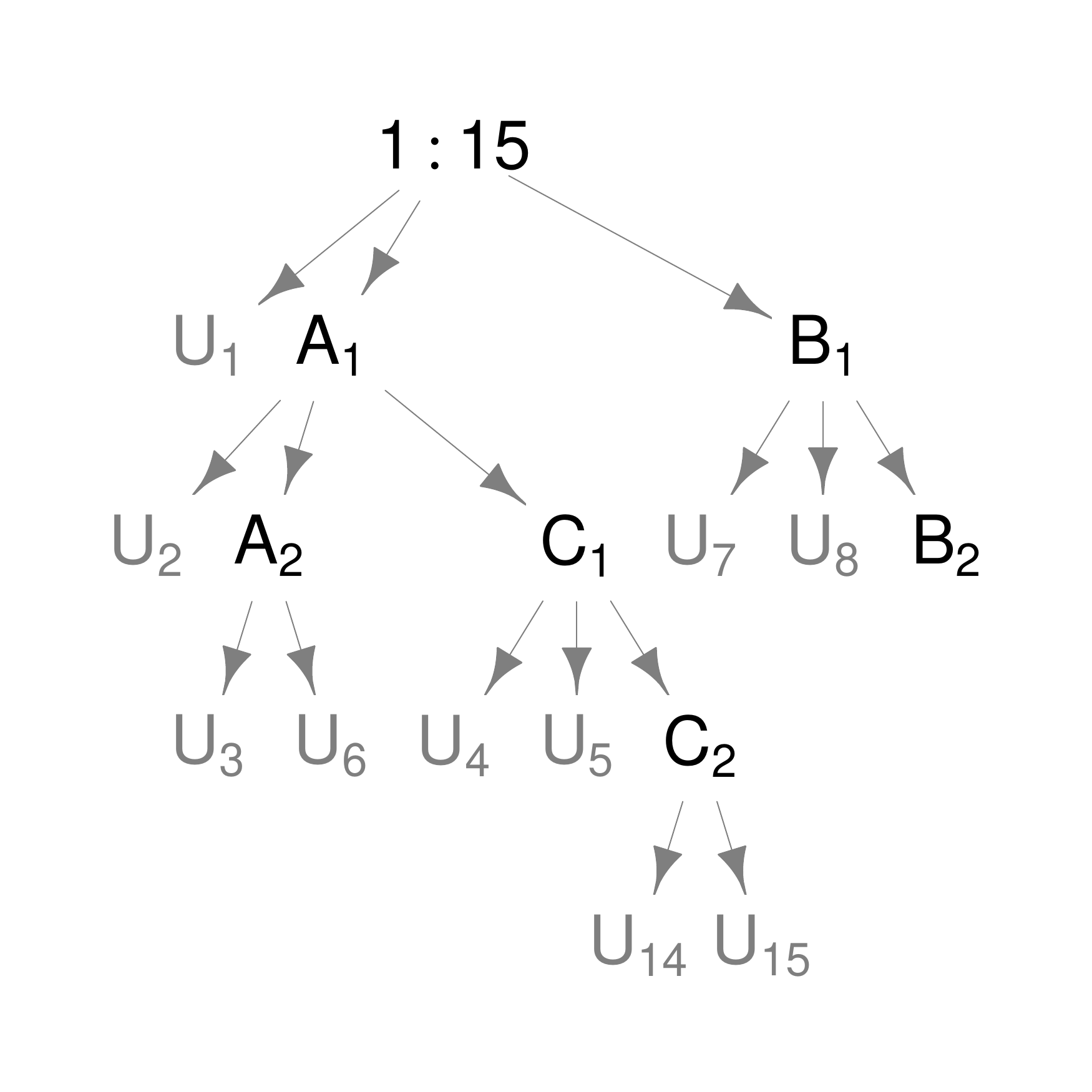}
\end{minipage}
\begin{minipage}[c]{0.4\textwidth}
\includegraphics[width=1\textwidth]{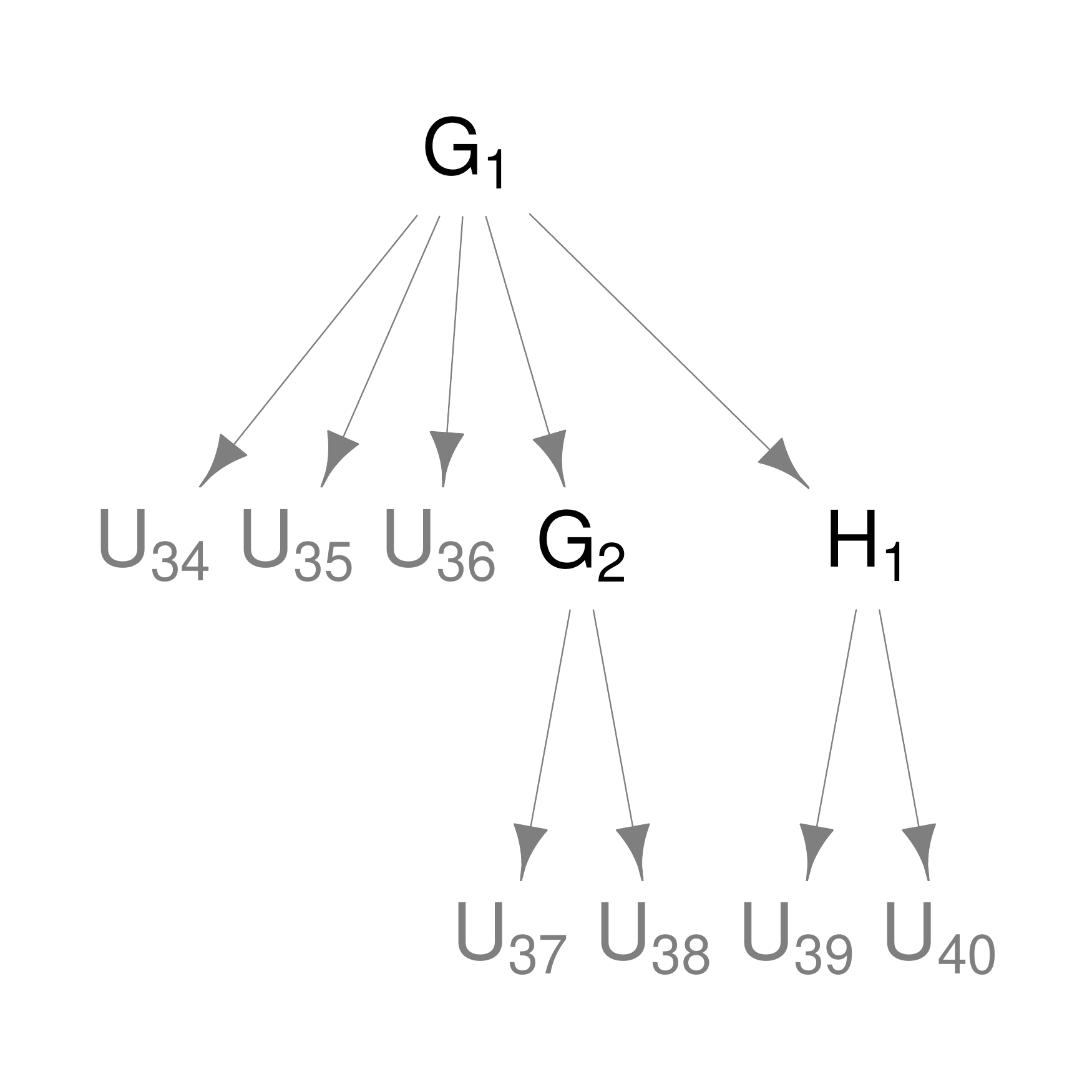}
\end{minipage}
\caption{Left: a fifteenvariate structure, where $B_2$ refer to an Archimedean copula spanned on $U_9$ through $U_{13}$. Right: the node $G_1$ from Figure \ref{large_struct}. \label{smaller_struct}}
\end{figure}

\begin{figure}[H]
\centering
\begin{minipage}[c]{1\textwidth}
\includegraphics[width=1\textwidth]{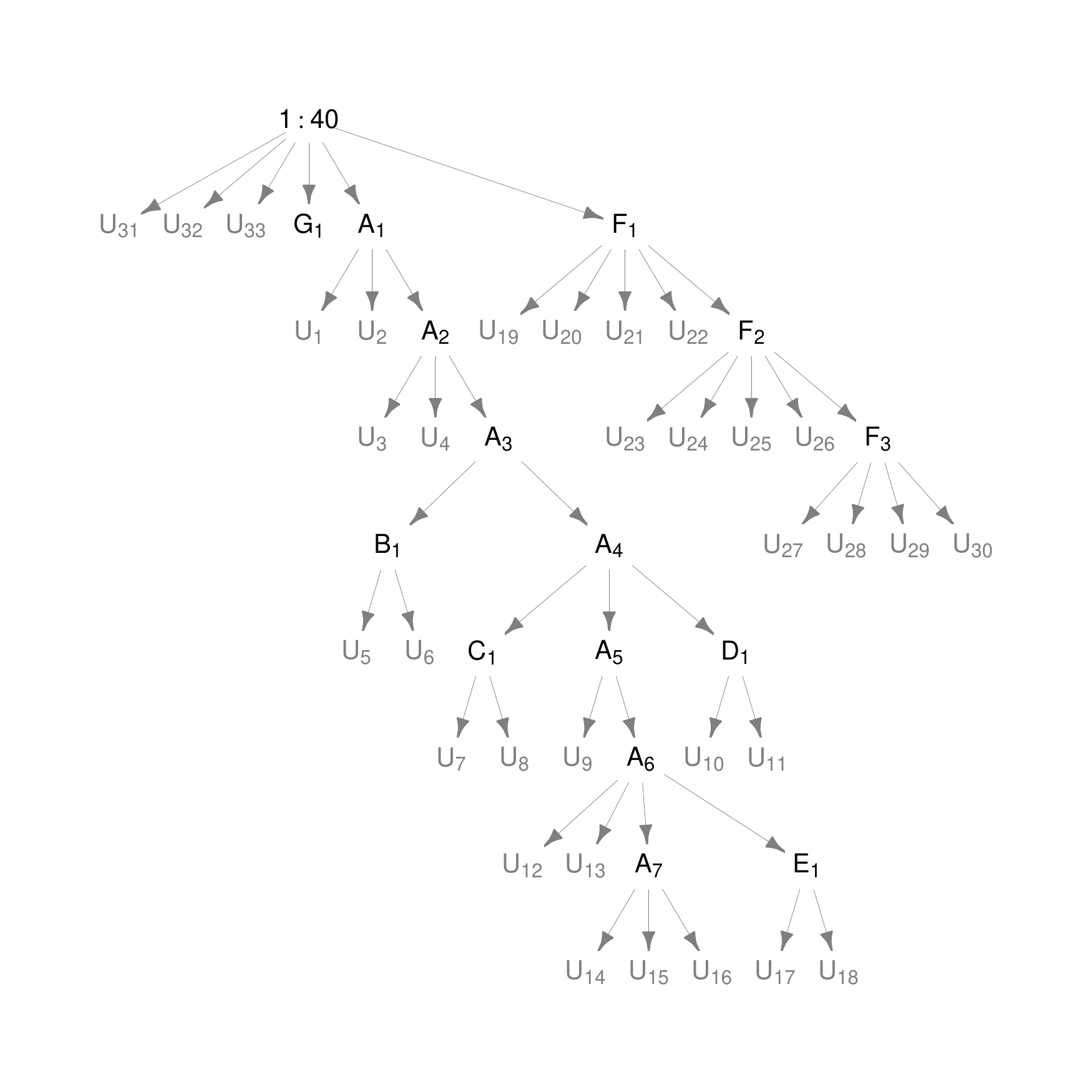}
\end{minipage}
\caption{A fortyvariate structure, the largest structure tested. Node $G_1$ is shown in Figure \ref{smaller_struct}.\label{large_struct}}
\end{figure}

Target structures are given in Figures \ref{small_struct}, \ref{smaller_struct} and \ref{large_struct}.

\begin{figure}[H]
\centering
\begin{tabular}{ccc}

\includegraphics[width=0.29\textwidth]{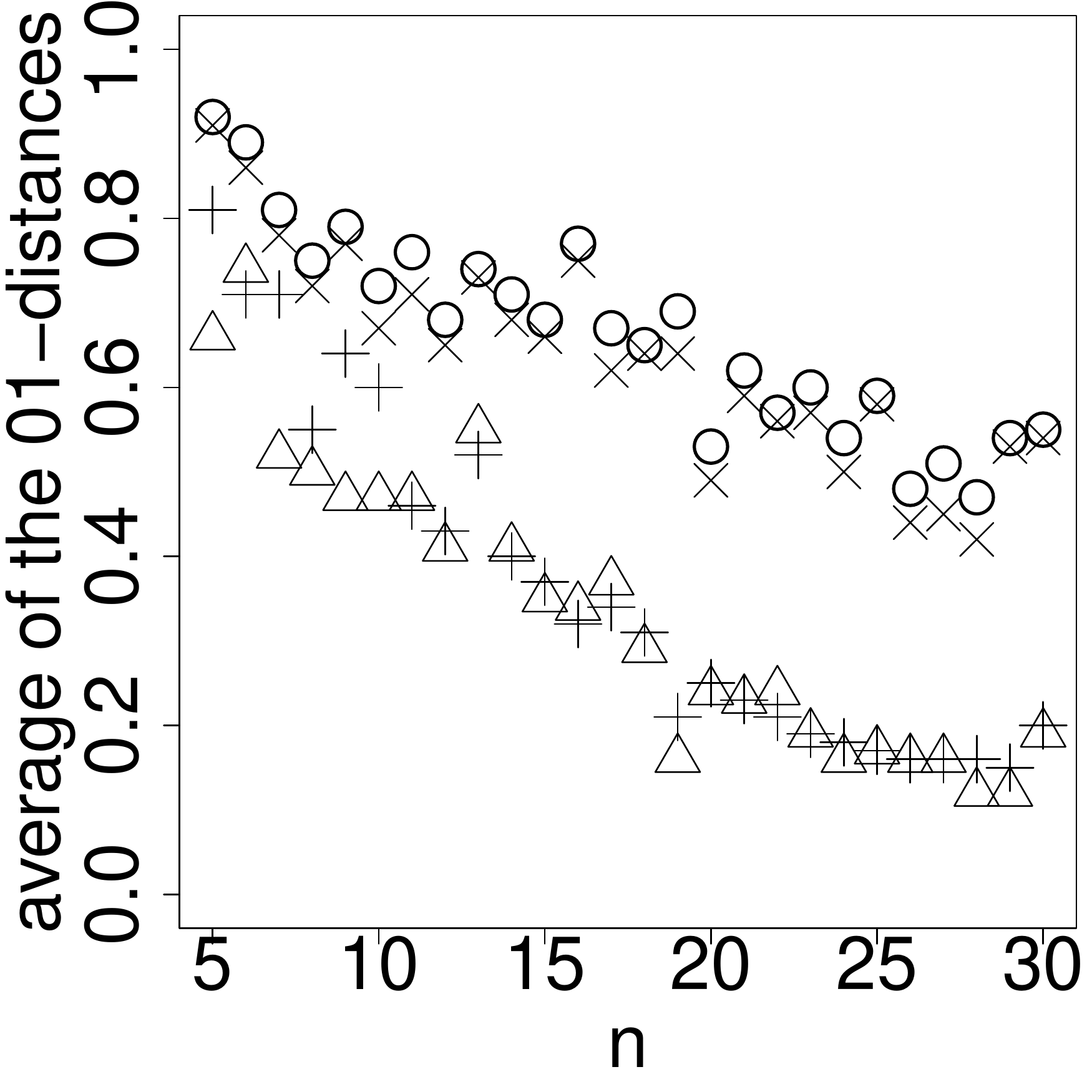}
&
\includegraphics[width=0.29\textwidth]{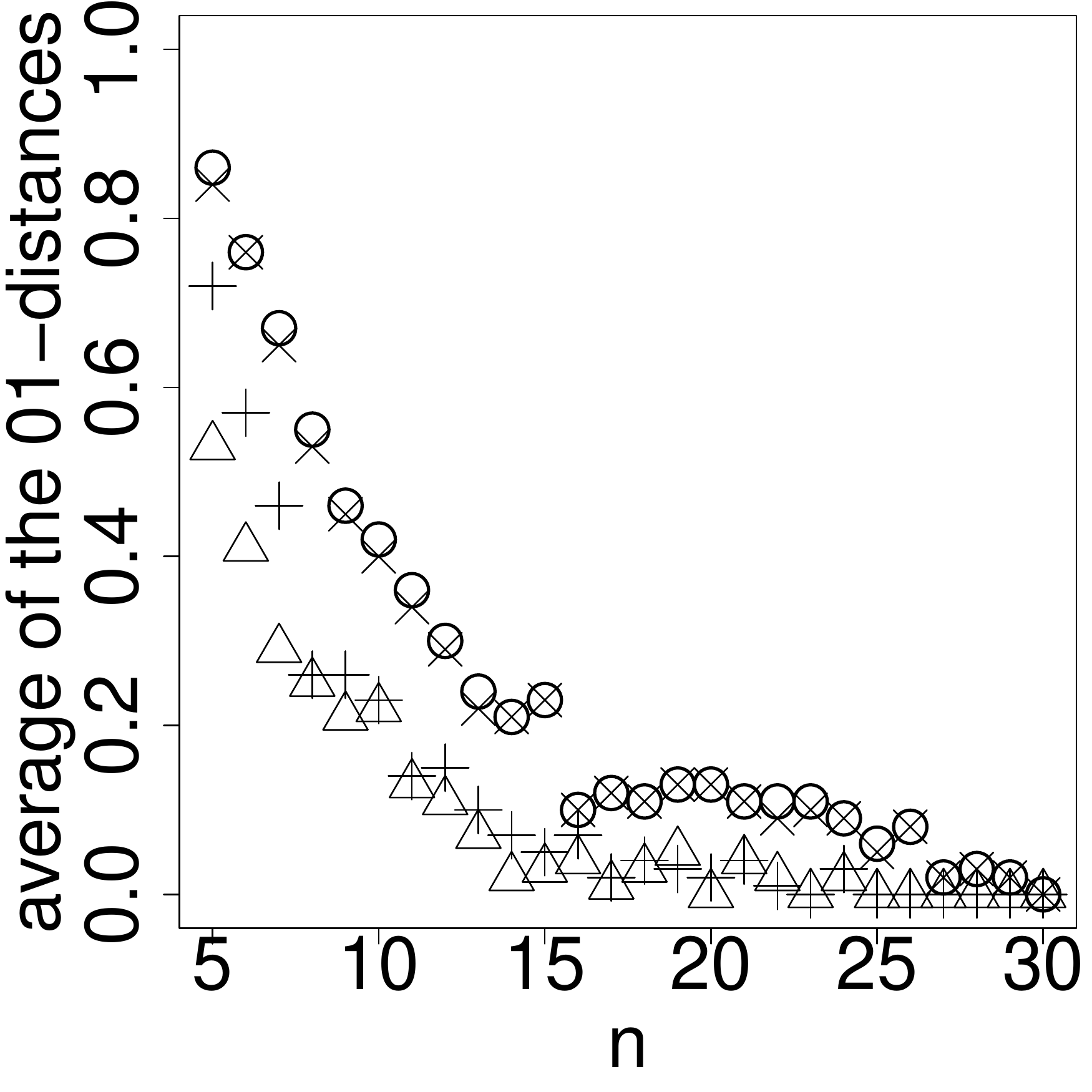}
&
\includegraphics[width=0.29\textwidth]{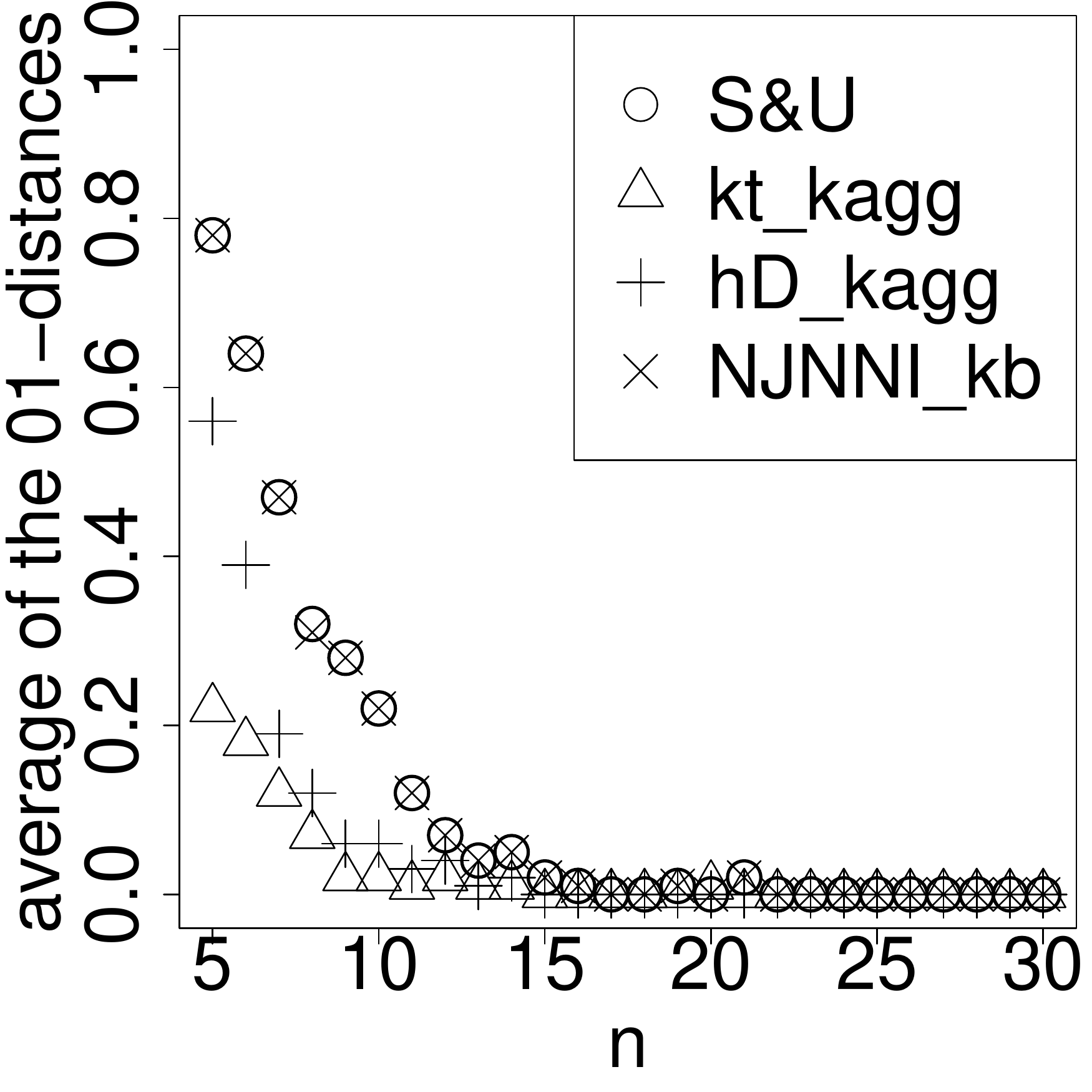}
\\
\includegraphics[width=0.29\textwidth]{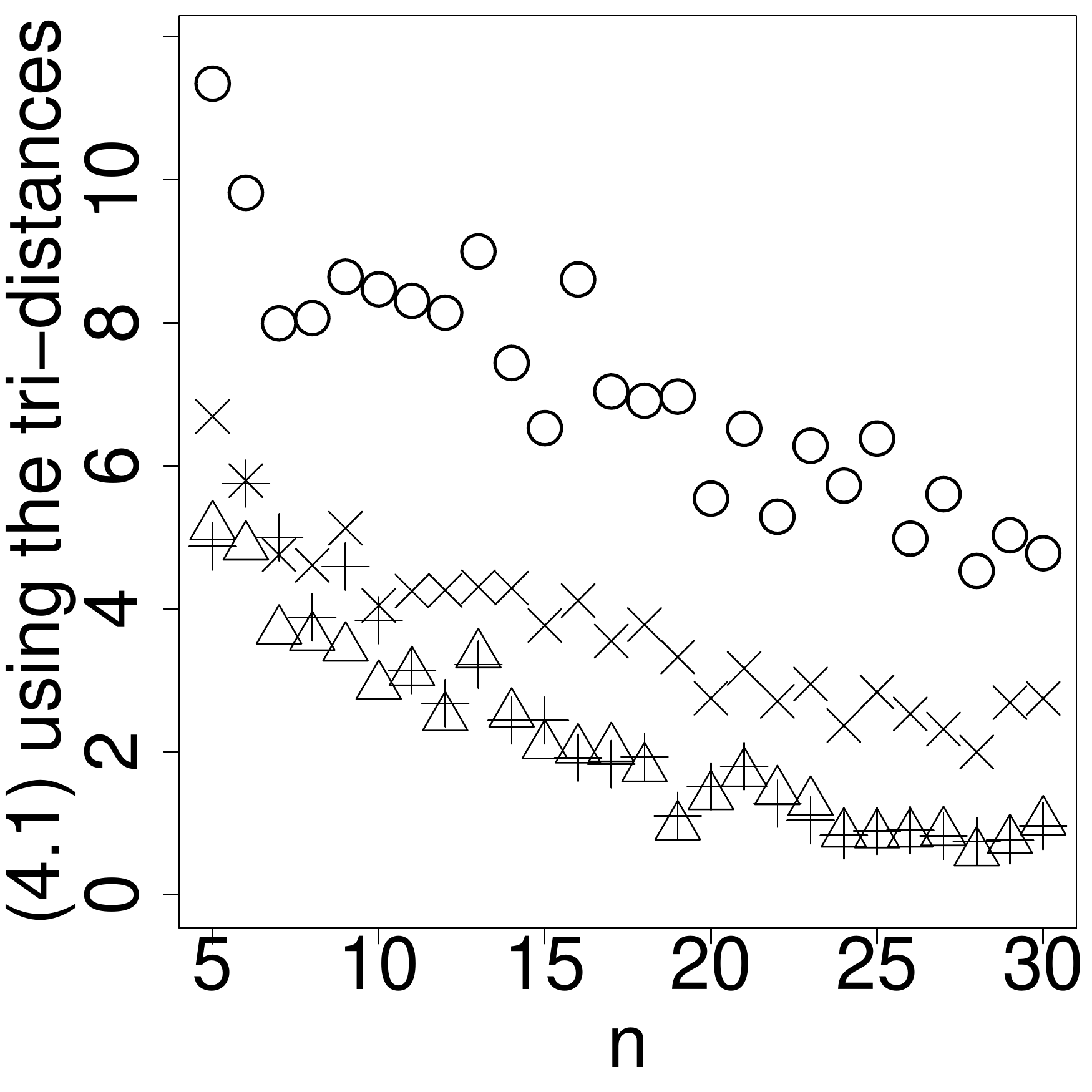}
&
\includegraphics[width=0.29\textwidth]{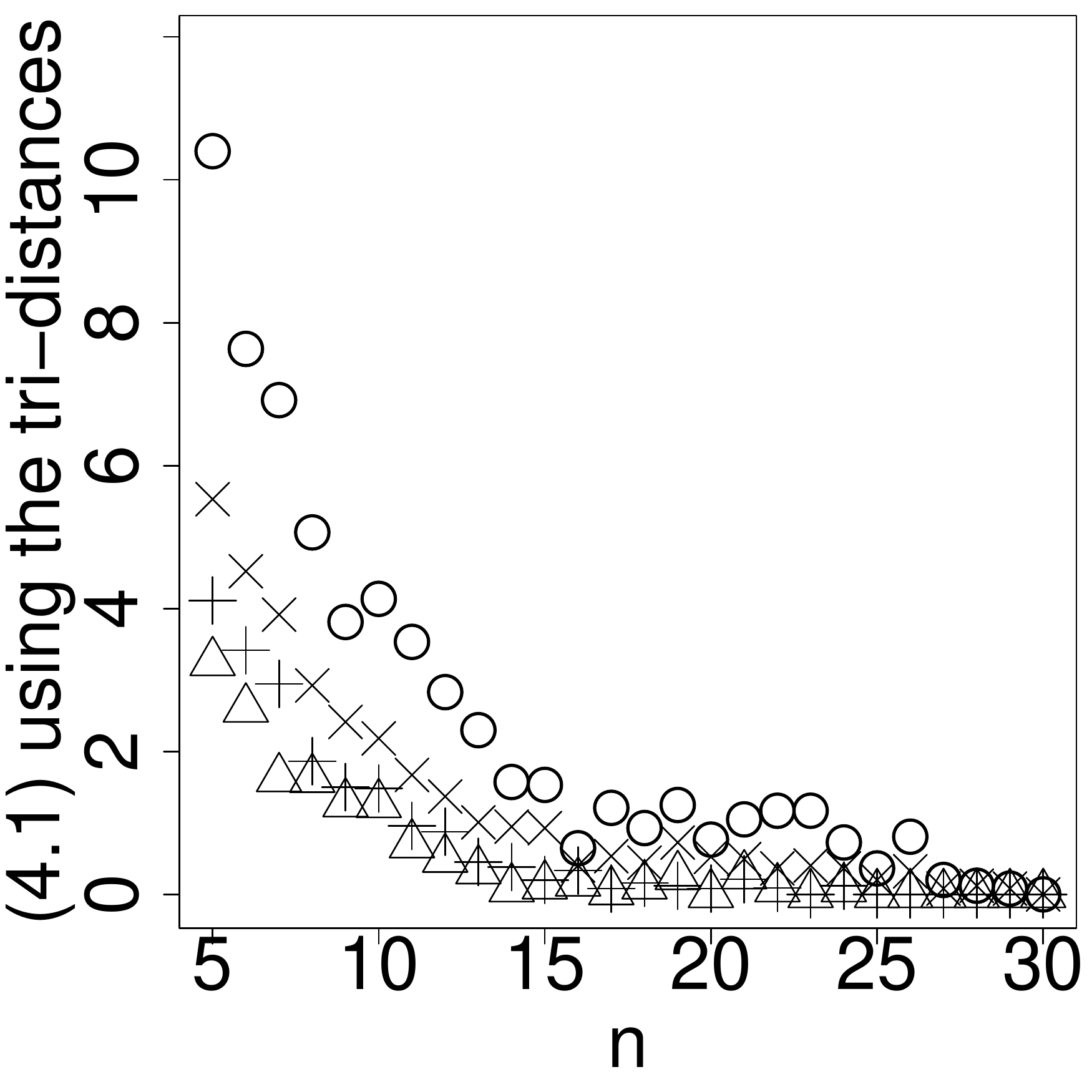}
&
\includegraphics[width=0.29\textwidth]{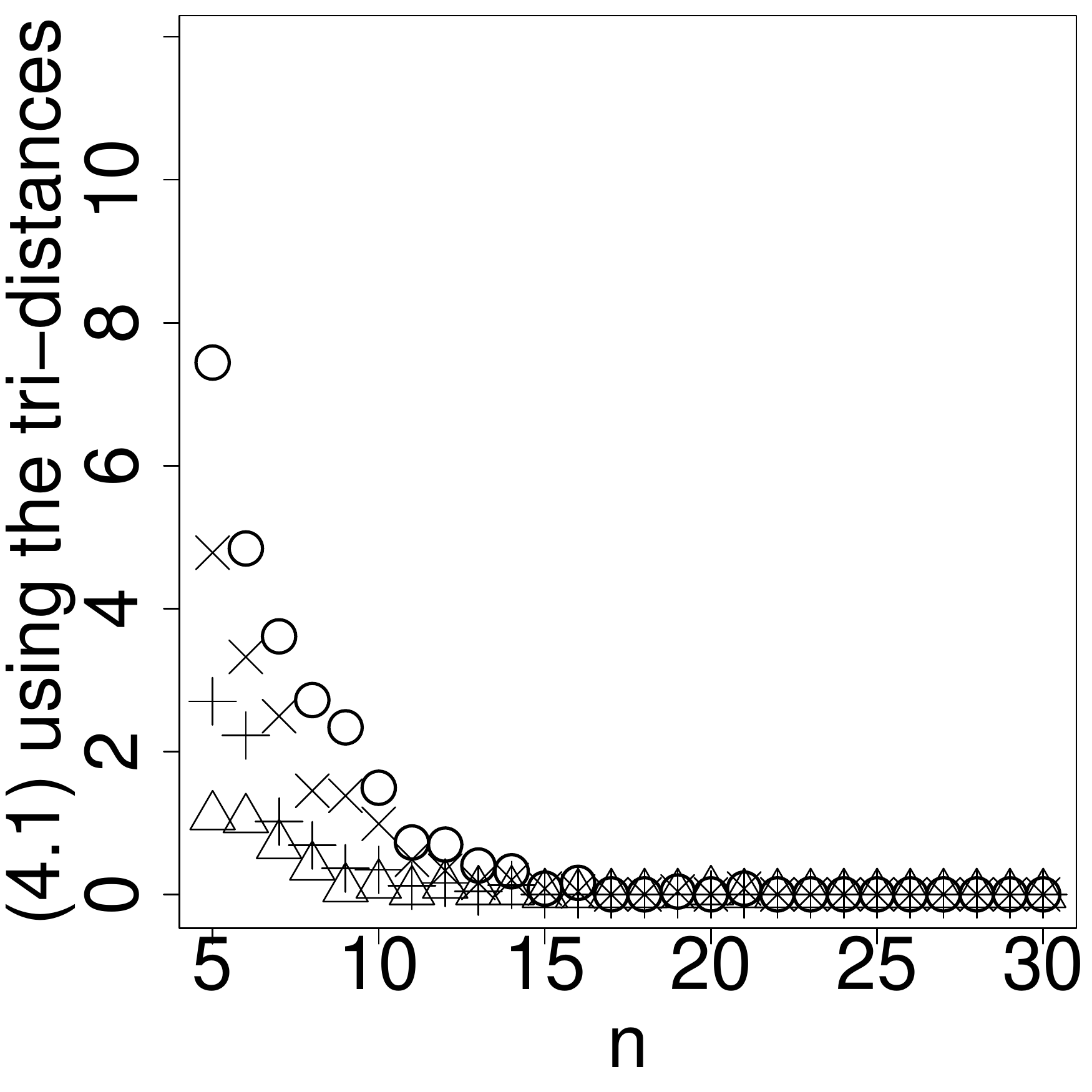}
\end{tabular}
\caption{results for the fourvariate binary structure. The generators used across the structure are all Clayton generators. The related sets of parameters are $(\tau_{1234}=0.4, \tau_{12}=0.6, \tau_{34}=0.6)$ for the left-hand side of the figure, $(\tau_{1234}=0.3, \tau_{12}=0.7, \tau_{34}=0.7)$ in the middle, and $(\tau_{1234}=0.2, \tau_{12}=0.8, \tau_{34}=0.8)$ for the right-hand side of the figure. \label{perf(12)(34)}}
\end{figure}

\begin{figure}[H]
\centering
\begin{tabular}{ccc}

\includegraphics[width=0.29\textwidth]{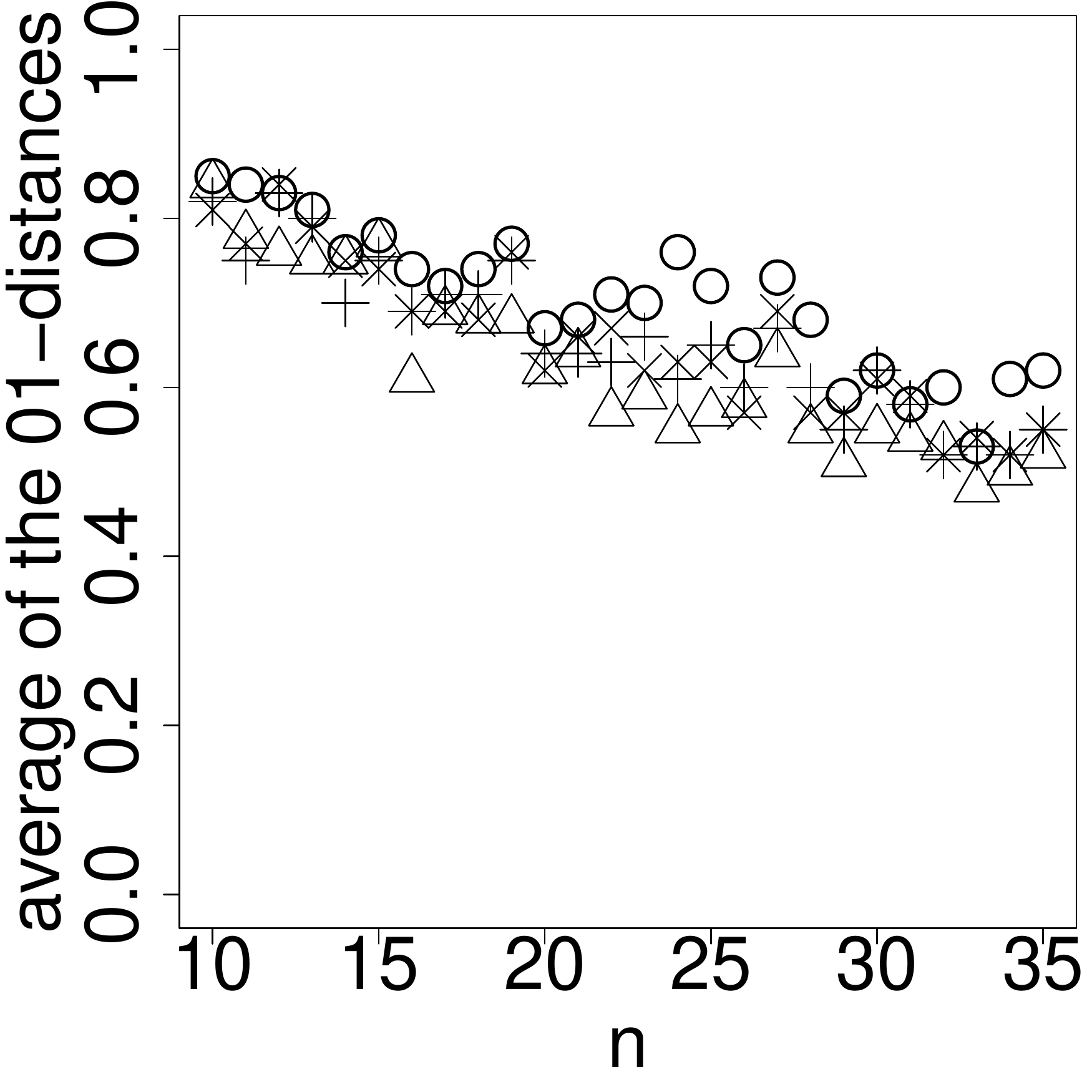}
&
\includegraphics[width=0.29\textwidth]{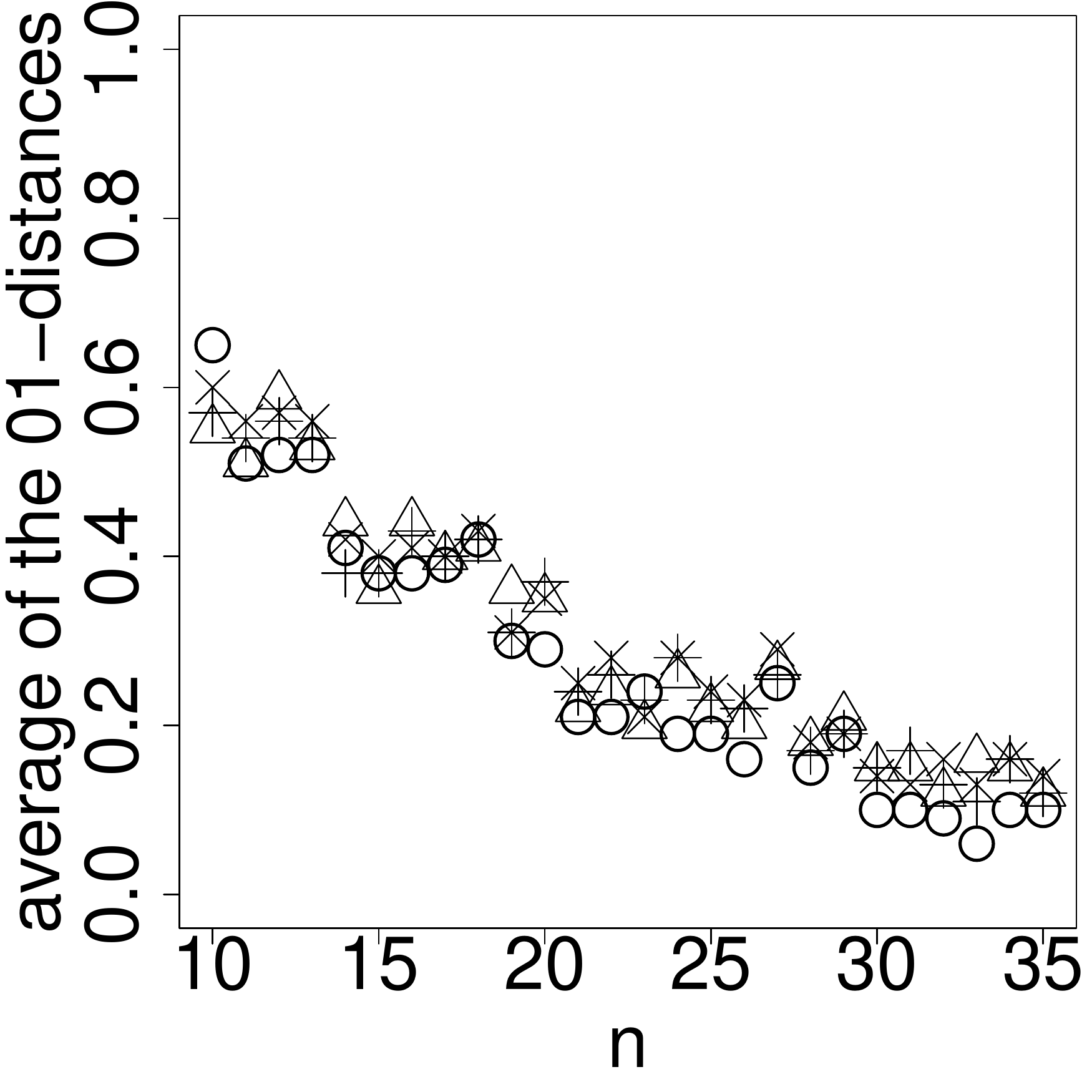}
&
\includegraphics[width=0.29\textwidth]{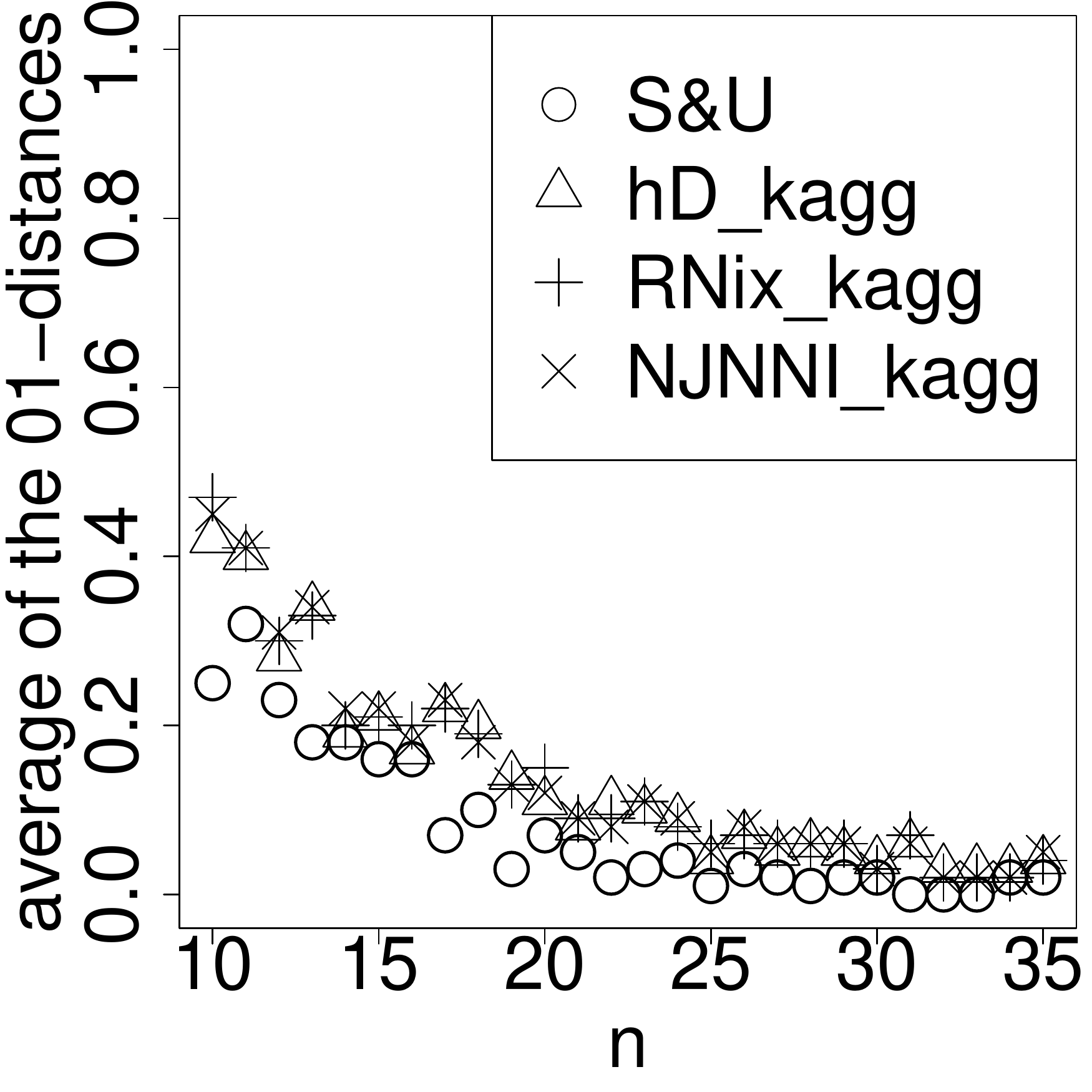}
\\
\includegraphics[width=0.29\textwidth]{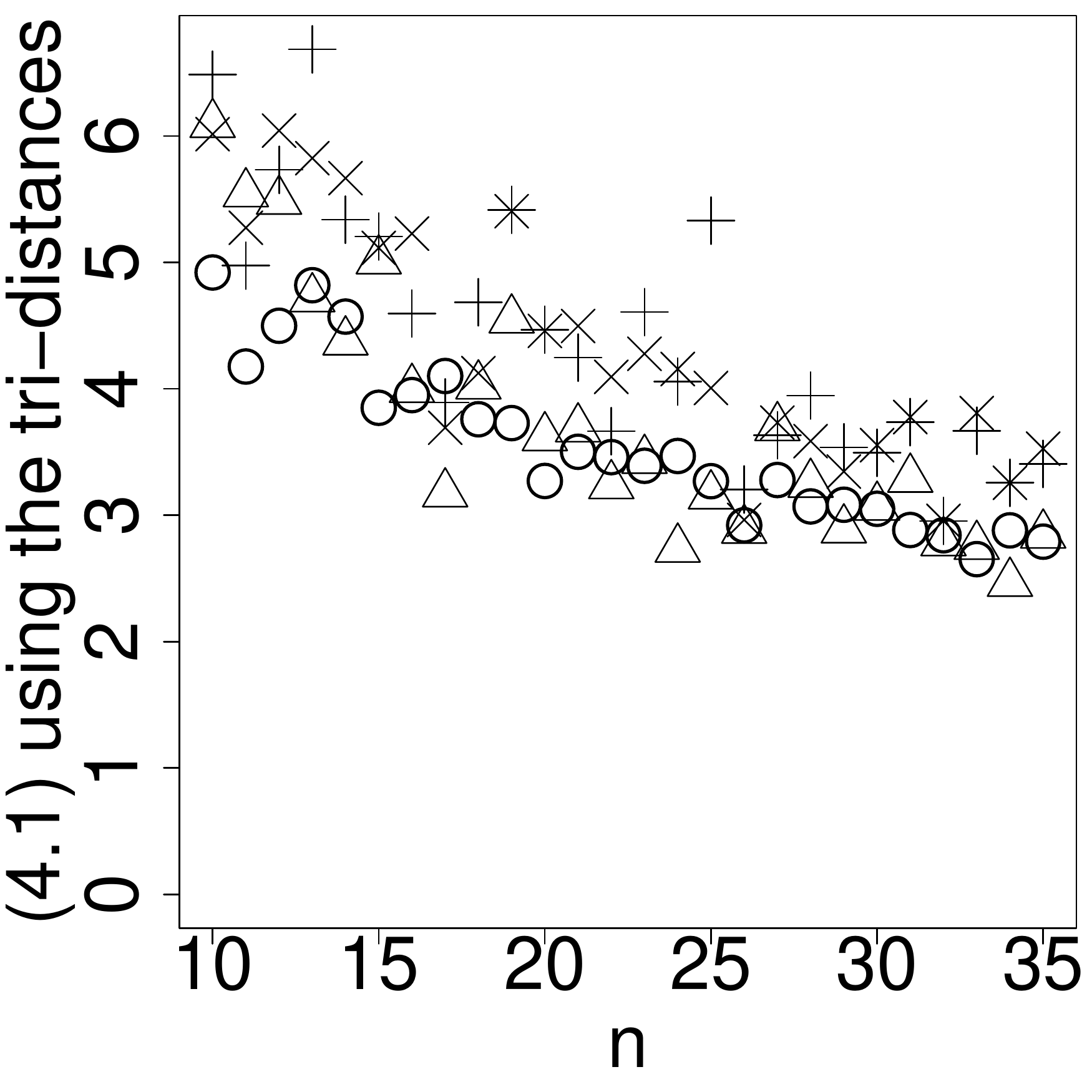}
&
\includegraphics[width=0.29\textwidth]{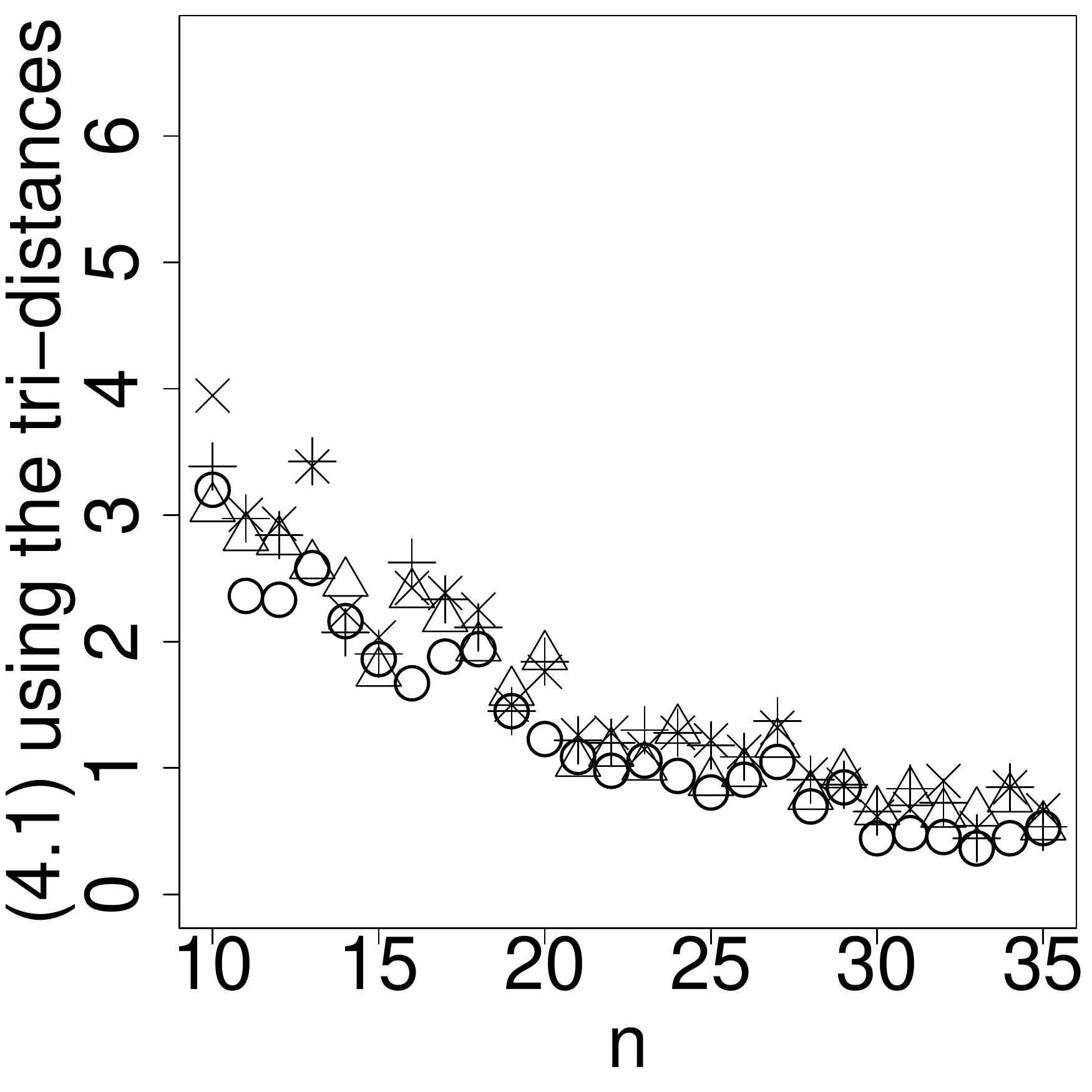}
&
\includegraphics[width=0.29\textwidth]{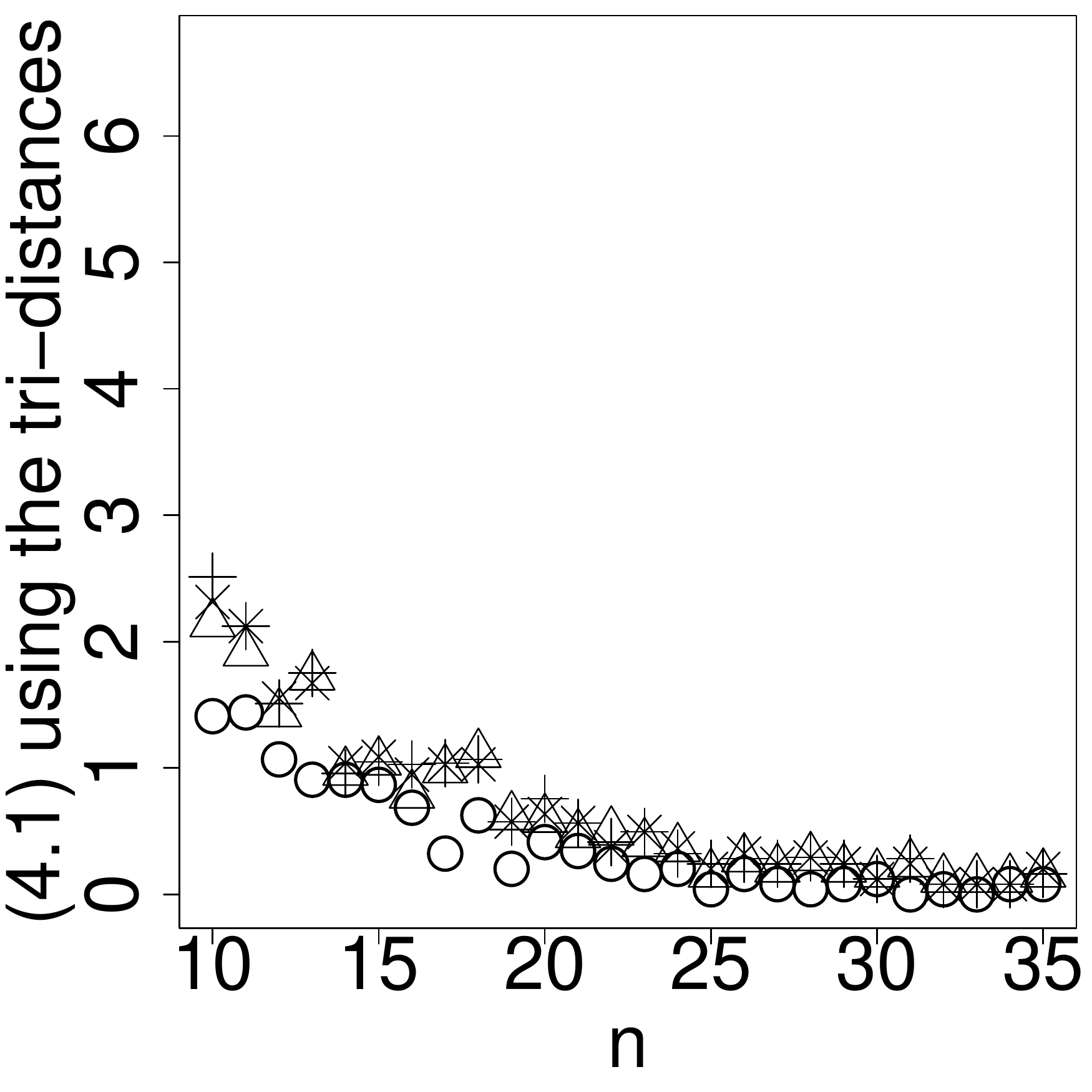}
\end{tabular}
\caption{results for the second fourvariate structure. The generators used across the structure are, again, all Clayton generators. The related sets of parameters are $(\tau_{1234}=0.4, \tau_{34}=0.6)$ for the hand-left side of the figure, $(\tau_{1234}=0.3, \tau_{34}=0.7)$ in the middle, and $(\tau_{1234}=0.2, \tau_{34}=0.8)$ for the hand-right side of the figure.\label{perf12(34)}}
\end{figure}

\begin{figure}[H]
\centering
\begin{tabular}{ccc}
\includegraphics[width=0.29\textwidth]{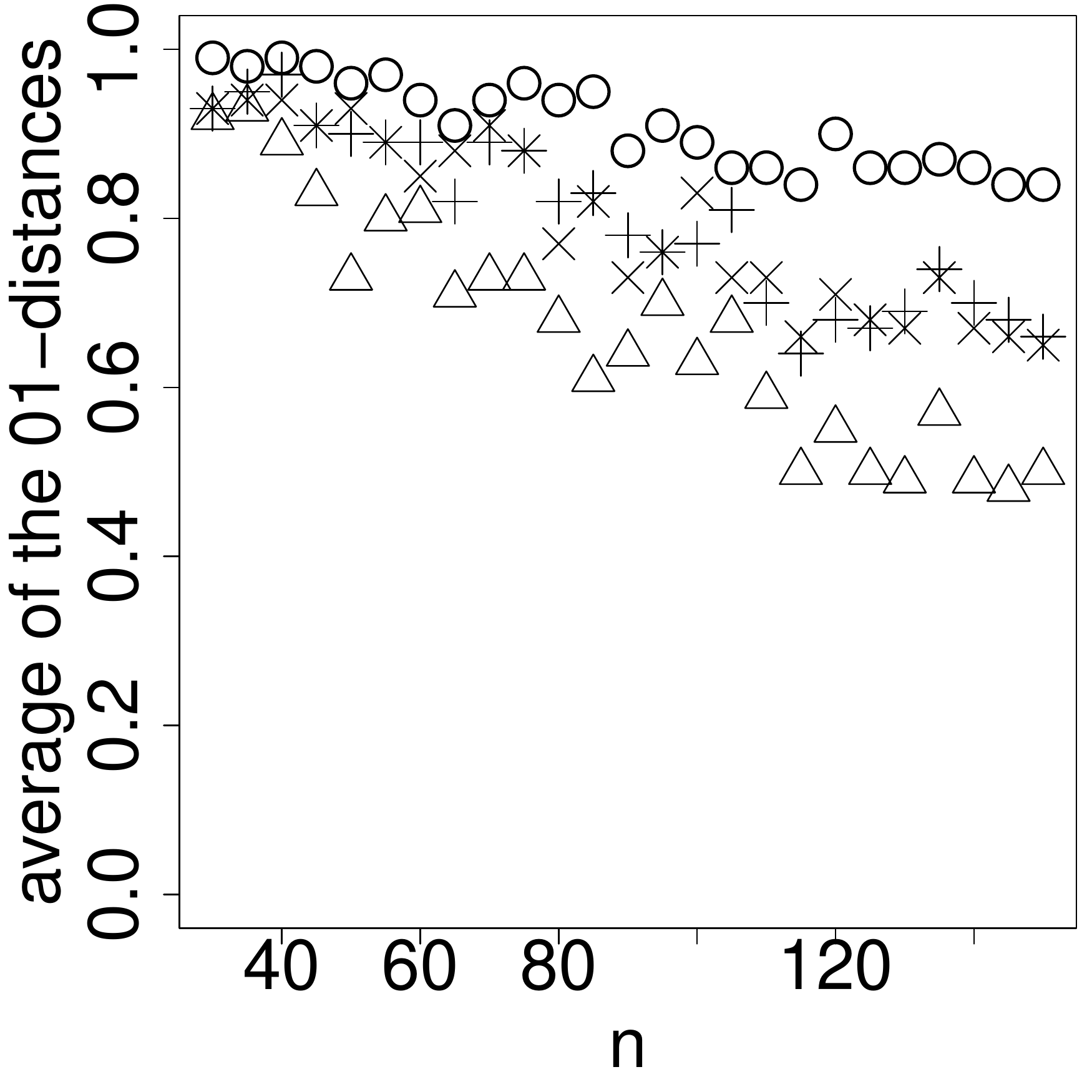}
&
\includegraphics[width=0.29\textwidth]{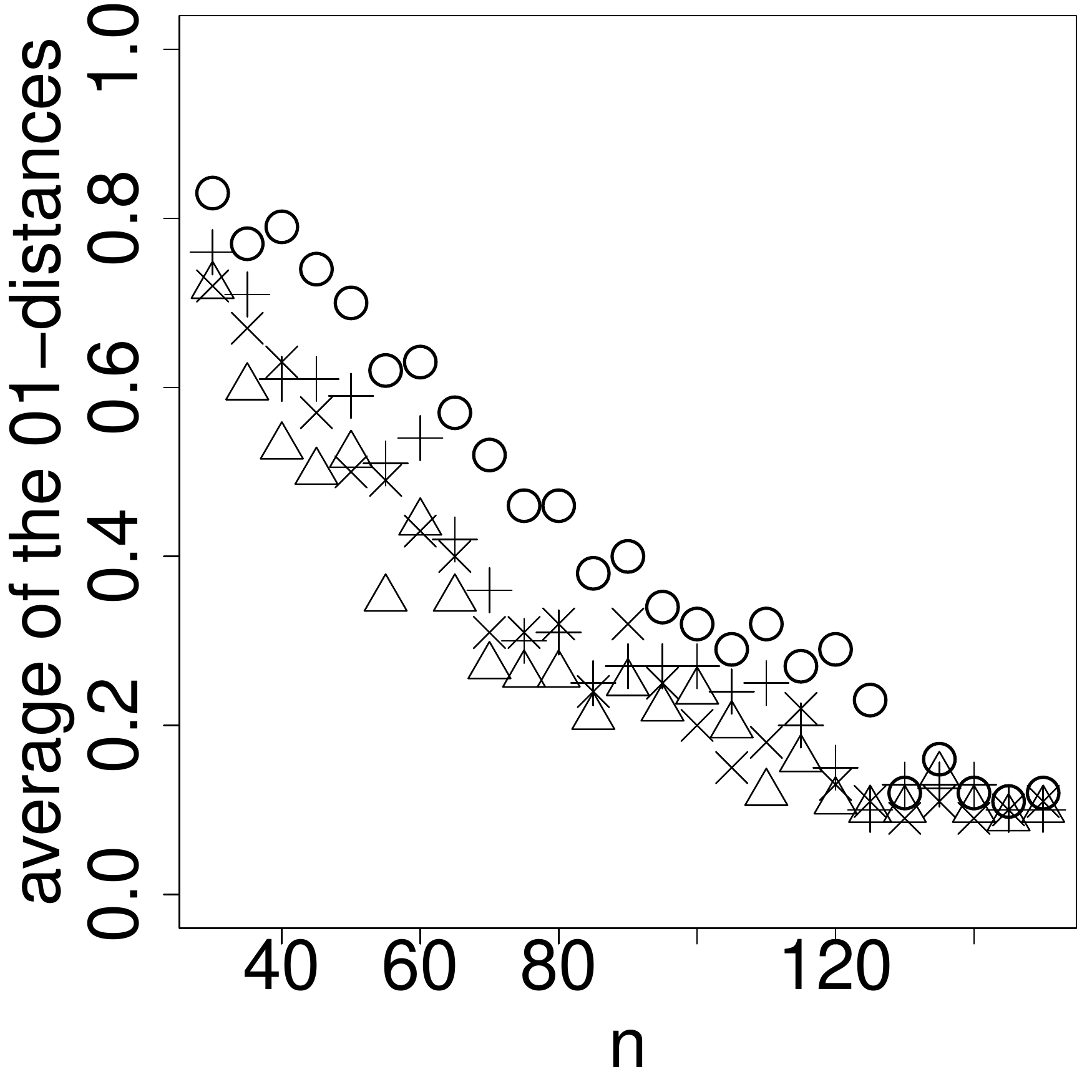}
&
\includegraphics[width=0.29\textwidth]{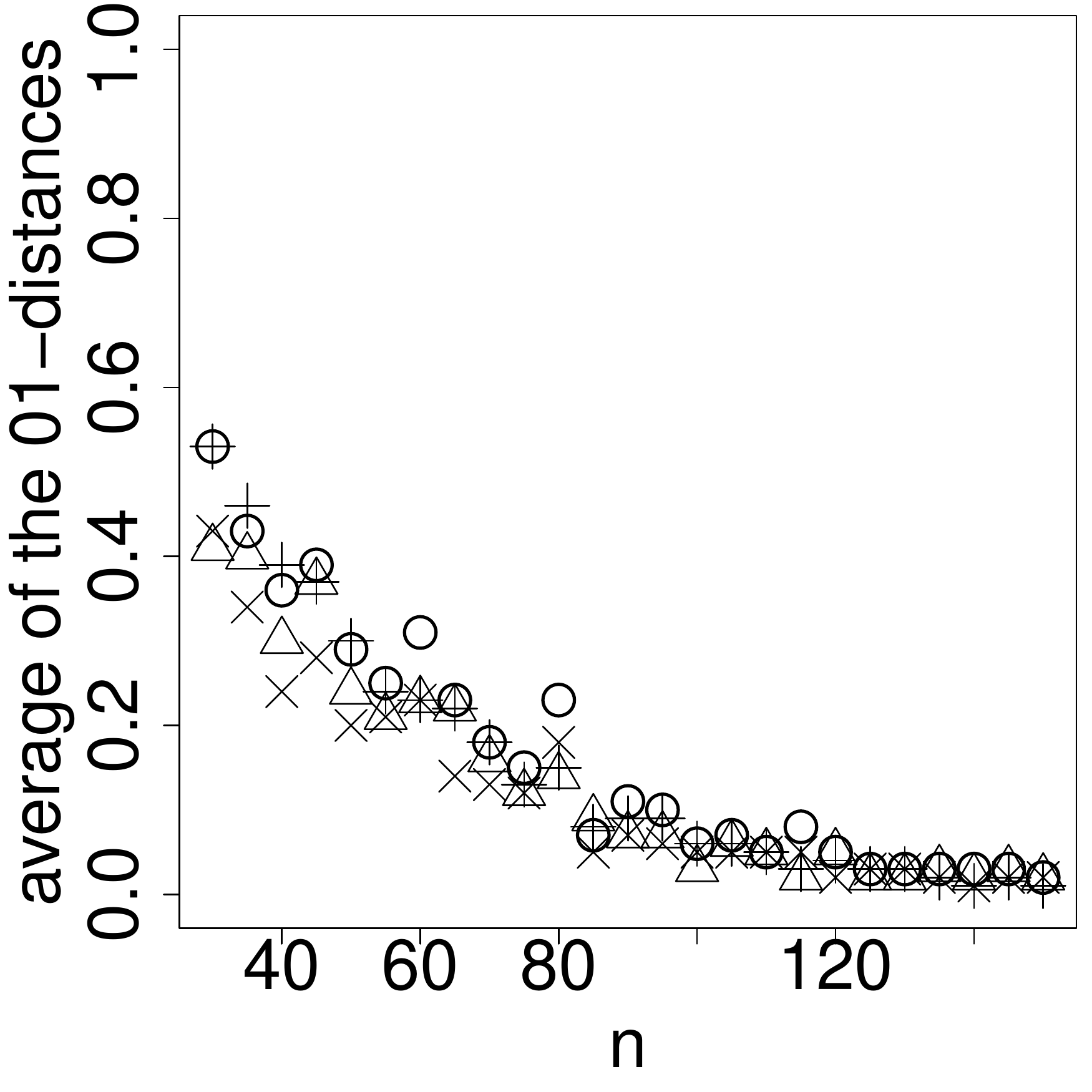}
\\
\includegraphics[width=0.29\textwidth]{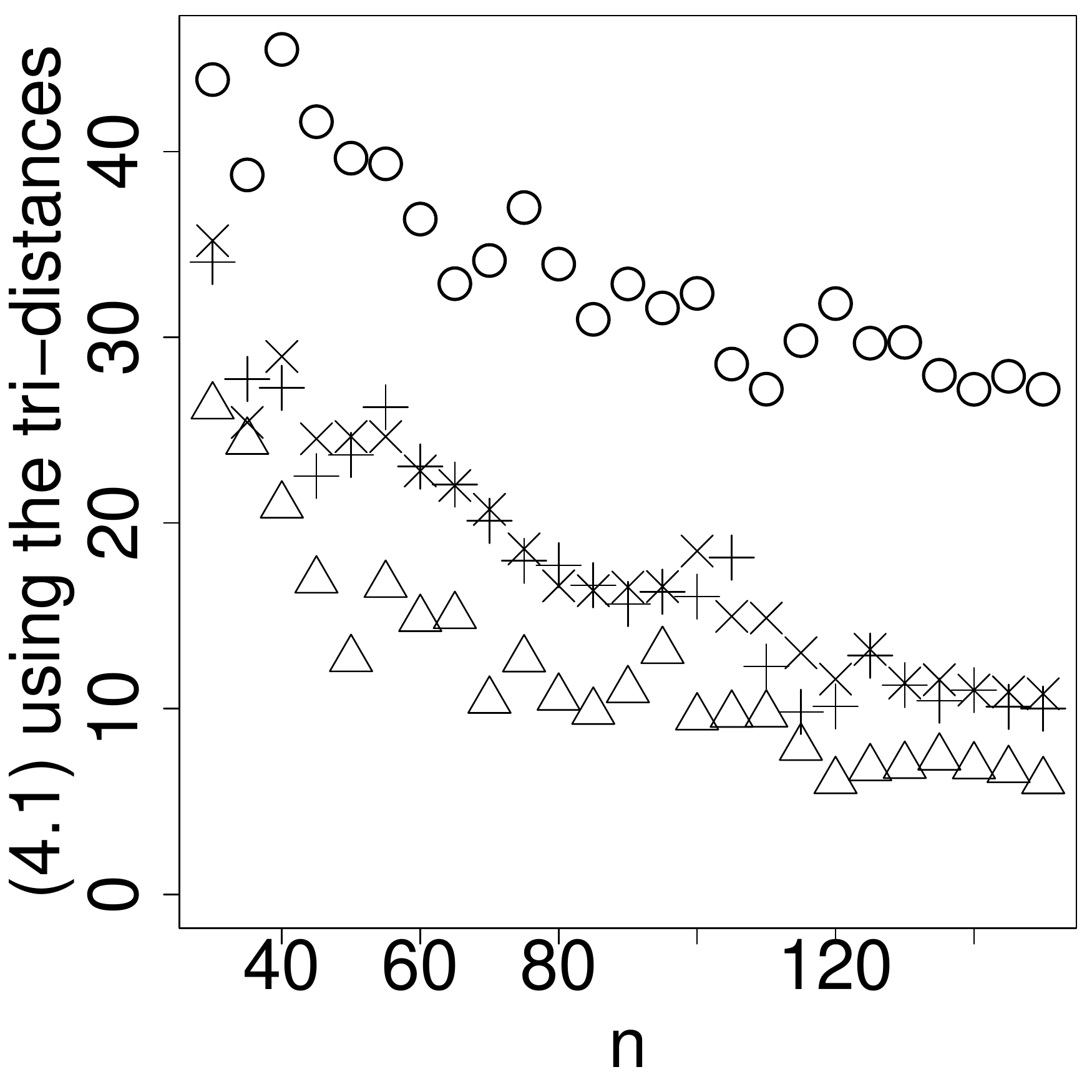}
&
\includegraphics[width=0.29\textwidth]{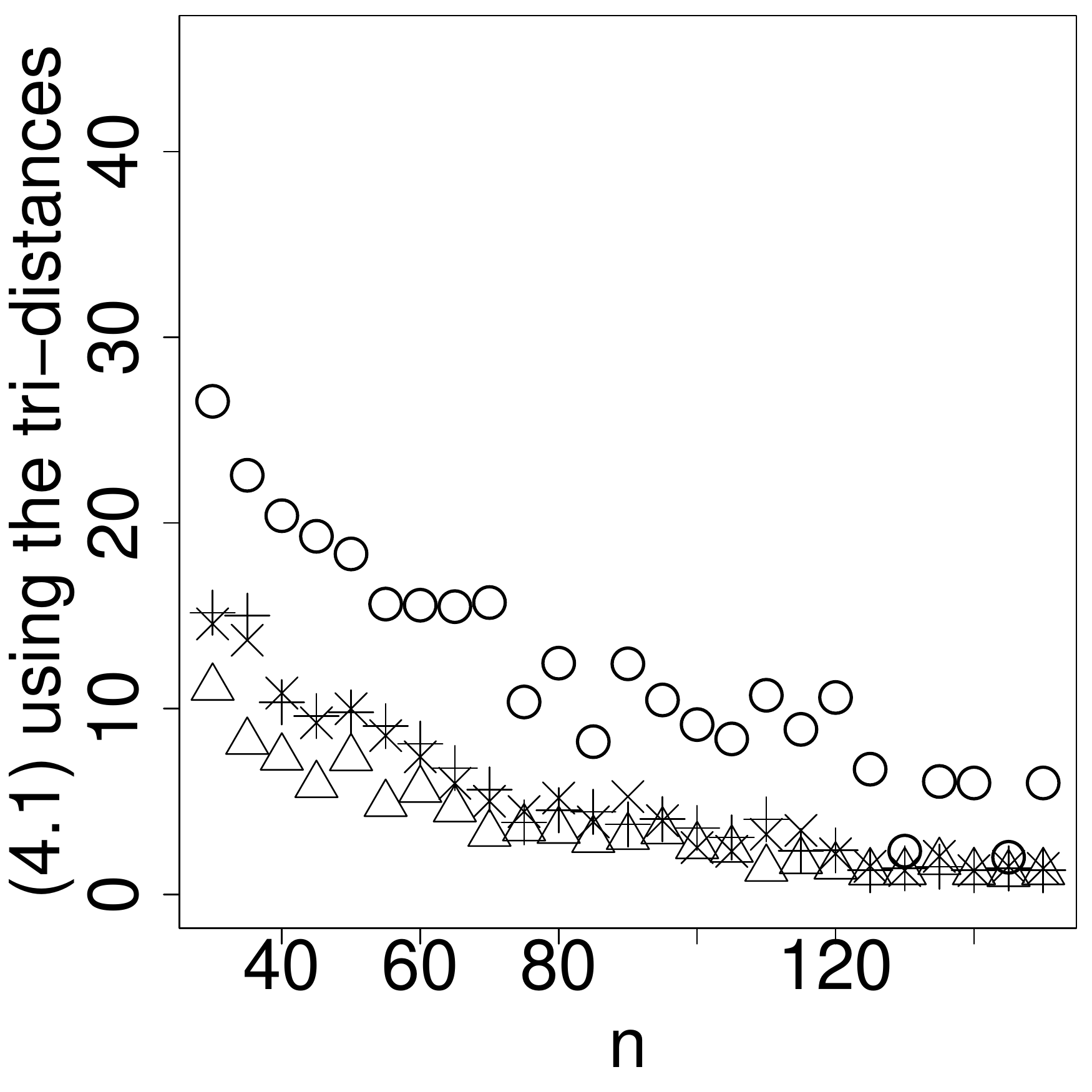}
&
\includegraphics[width=0.29\textwidth]{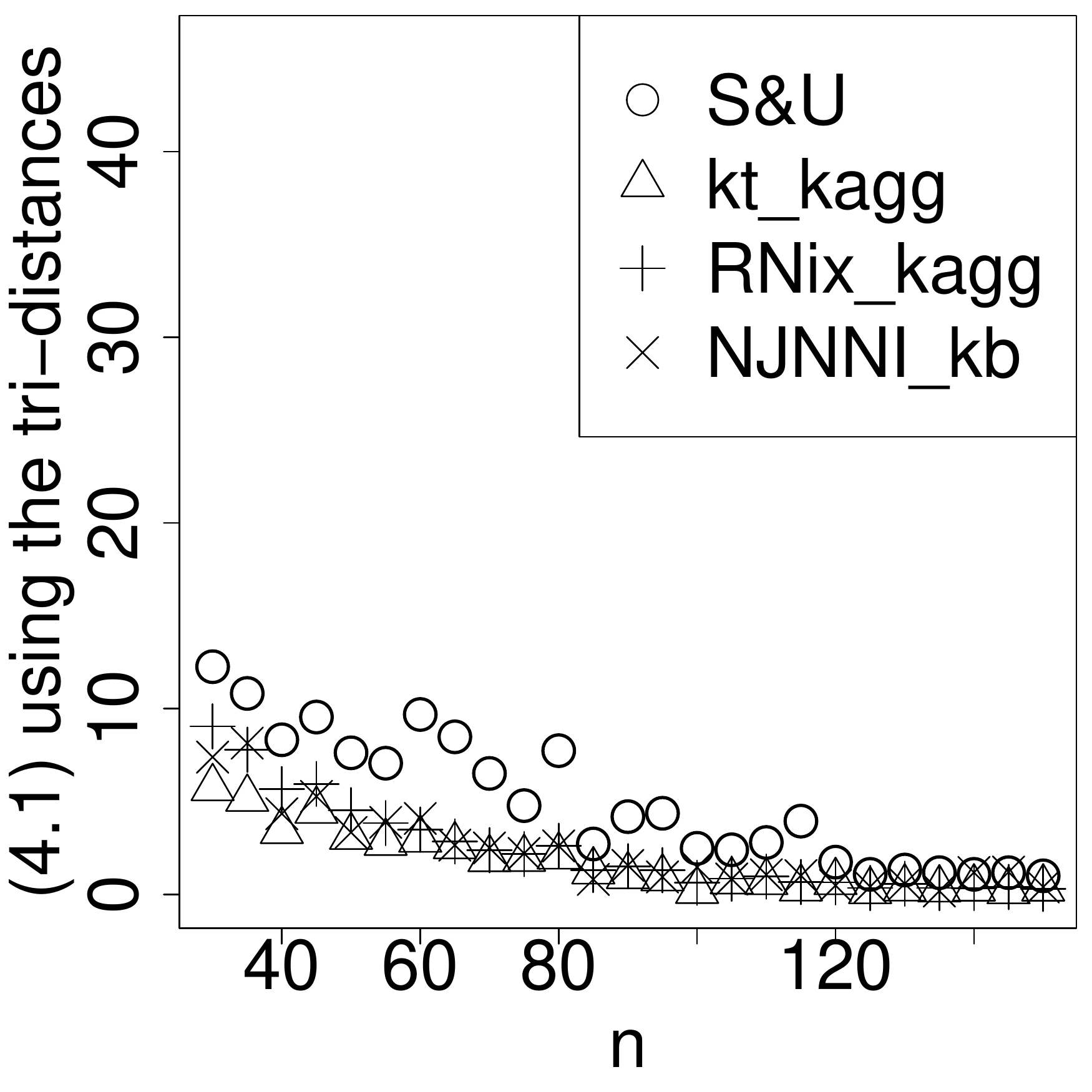}
\end{tabular}
\caption{results for the fivevariate structure. The generators used across the structure are all Gumbel generators. The related sets of parameters are $(\tau_{1:5}=0.4, \tau_{345}=0.5, \tau_{34}=0.6)$ for the left-hand side of the figure, $(\tau_{1:5}=0.3, \tau_{345}=0.5, \tau_{34}=0.7)$ in the middle, and $(\tau_{1:5}=0.2, \tau_{345}=0.5, \tau_{34}=0.8)$ for the right-hand side of the figure.\label{perffive}}
\end{figure}

\begin{figure}[H]
\centering
\begin{tabular}{cc}
\includegraphics[width=0.29\textwidth]{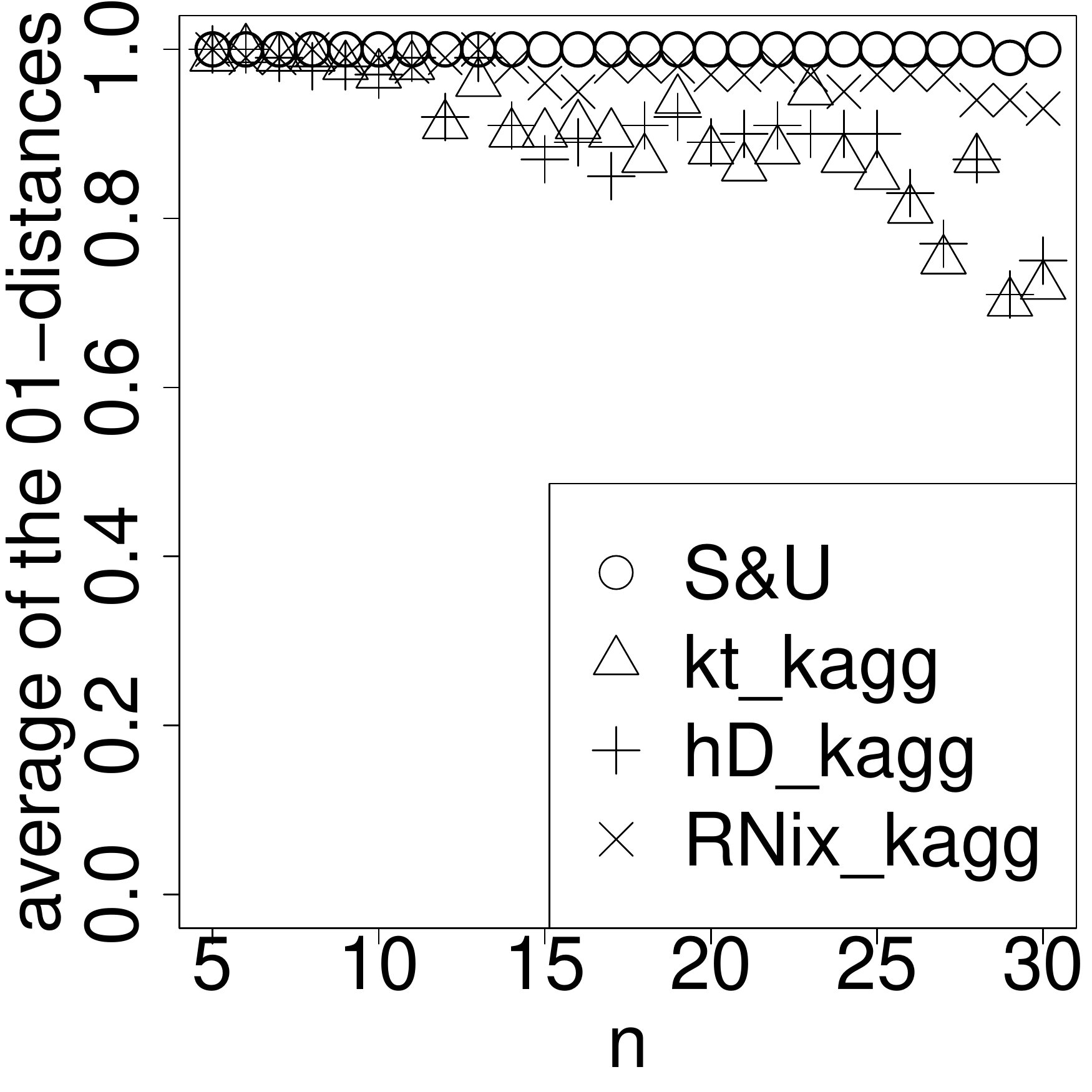}
&
\includegraphics[width=0.29\textwidth]{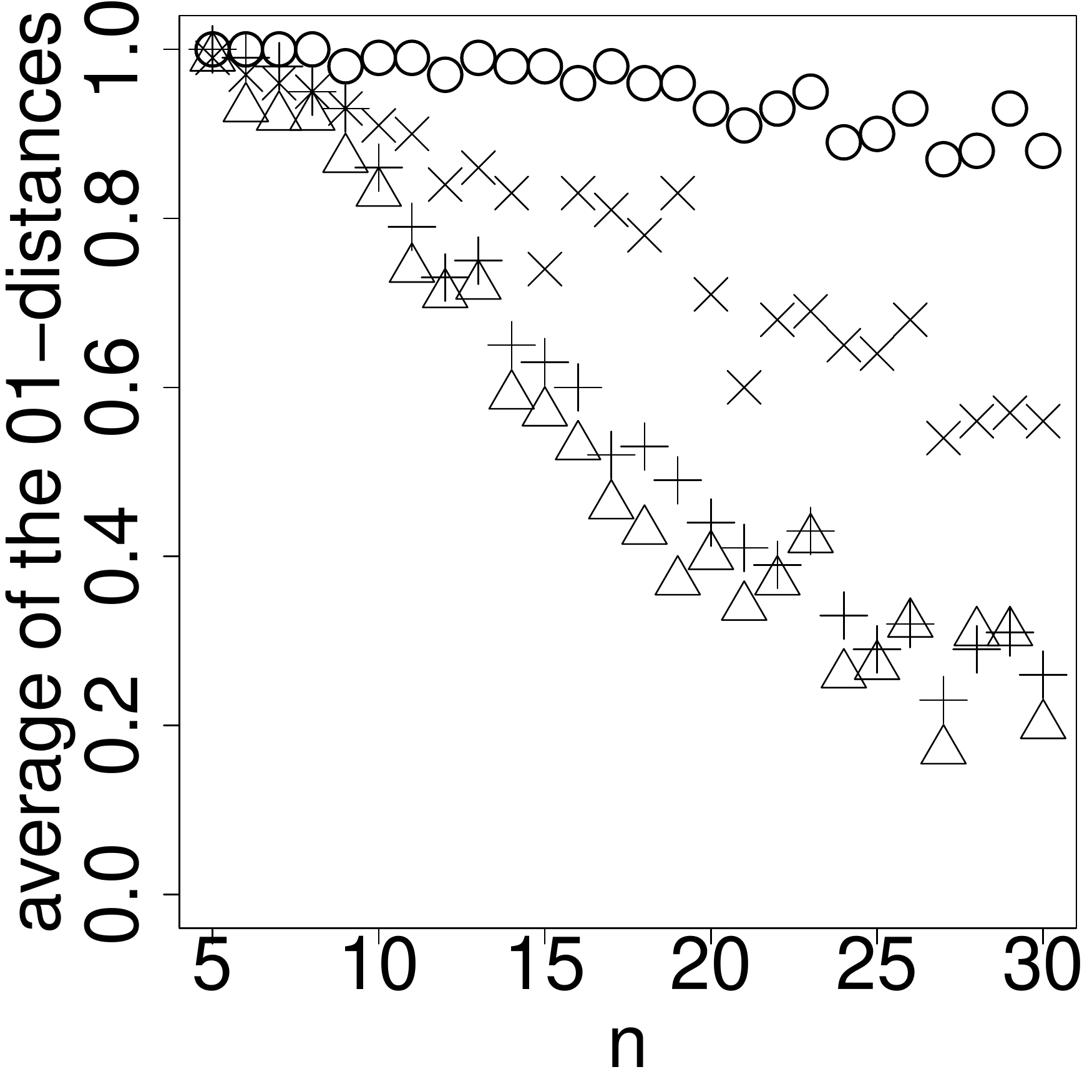}

\\
\includegraphics[width=0.29\textwidth]{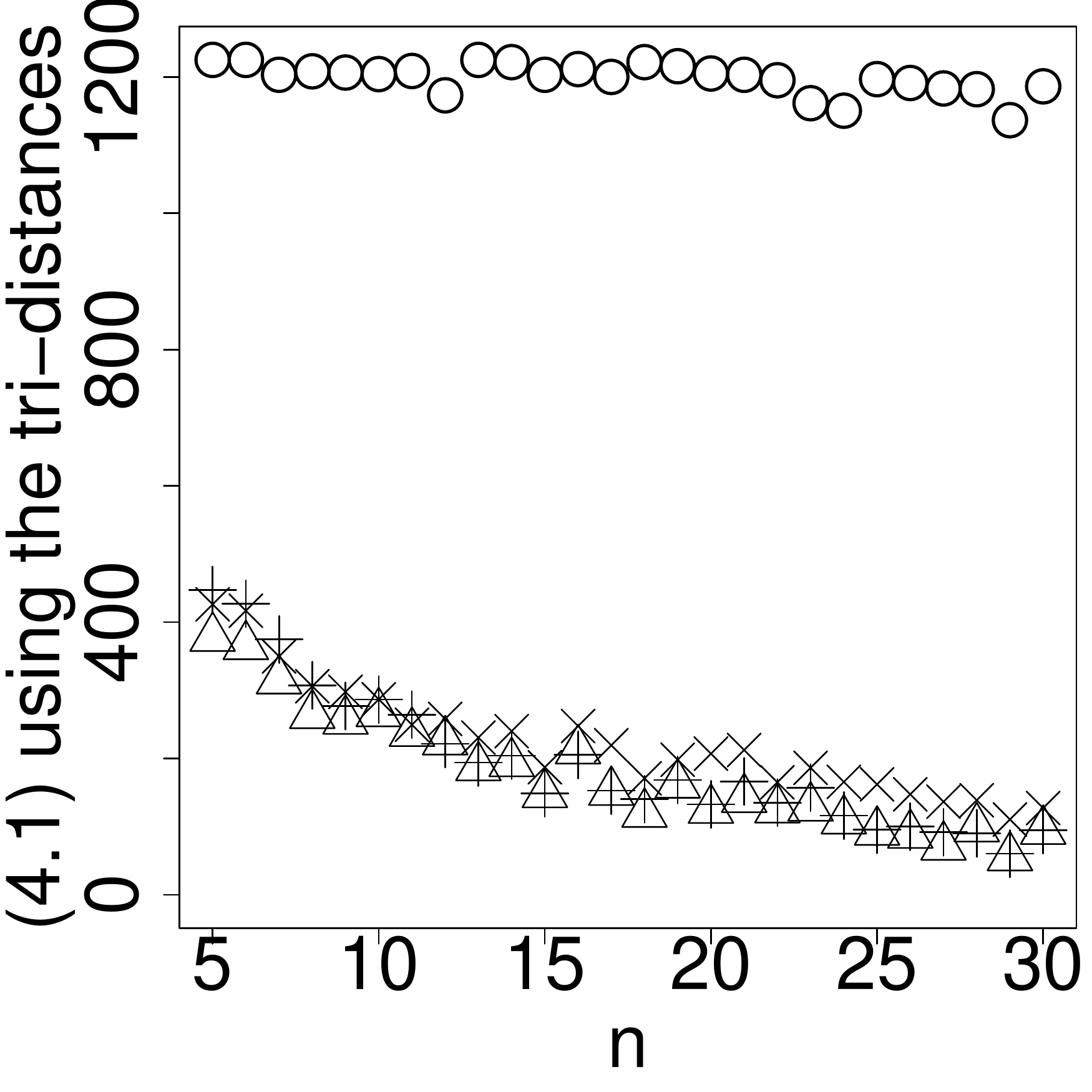}
&
\includegraphics[width=0.29\textwidth]{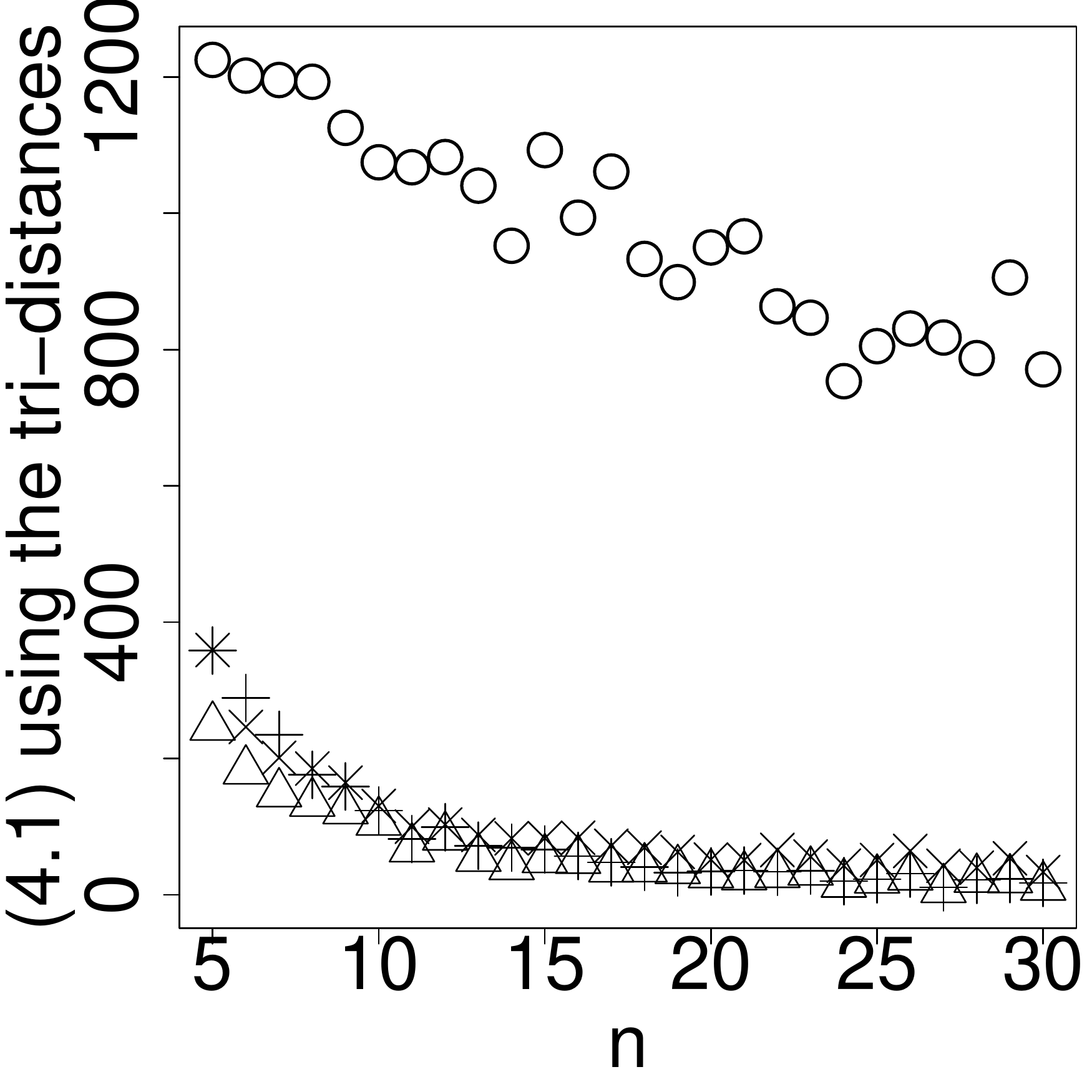}

\end{tabular}
\caption{results for the sevenvariate structure. The generators used across the structure are all Frank generators. The related sets of parameters are $(\tau_{1:7}=0.35, \tau_{123}=0.5, \tau_{23}=0.65, \tau_{4:7}=0.45, \tau_{567}=0.55, \tau_{67}=0.65)$ for the left-hand side of the figure and $(\tau_{1:7}=0.2, \tau_{123}=0.5, \tau_{23}=0.8, \tau_{4:7}=0.4, \tau_{567}=0.6, \tau_{67}=0.8)$ for the right-hand side of the figure.
 \label{perfseven}}
\end{figure}

\begin{figure}[H]
\centering
\begin{tabular}{cc}
\includegraphics[width=0.29\textwidth]{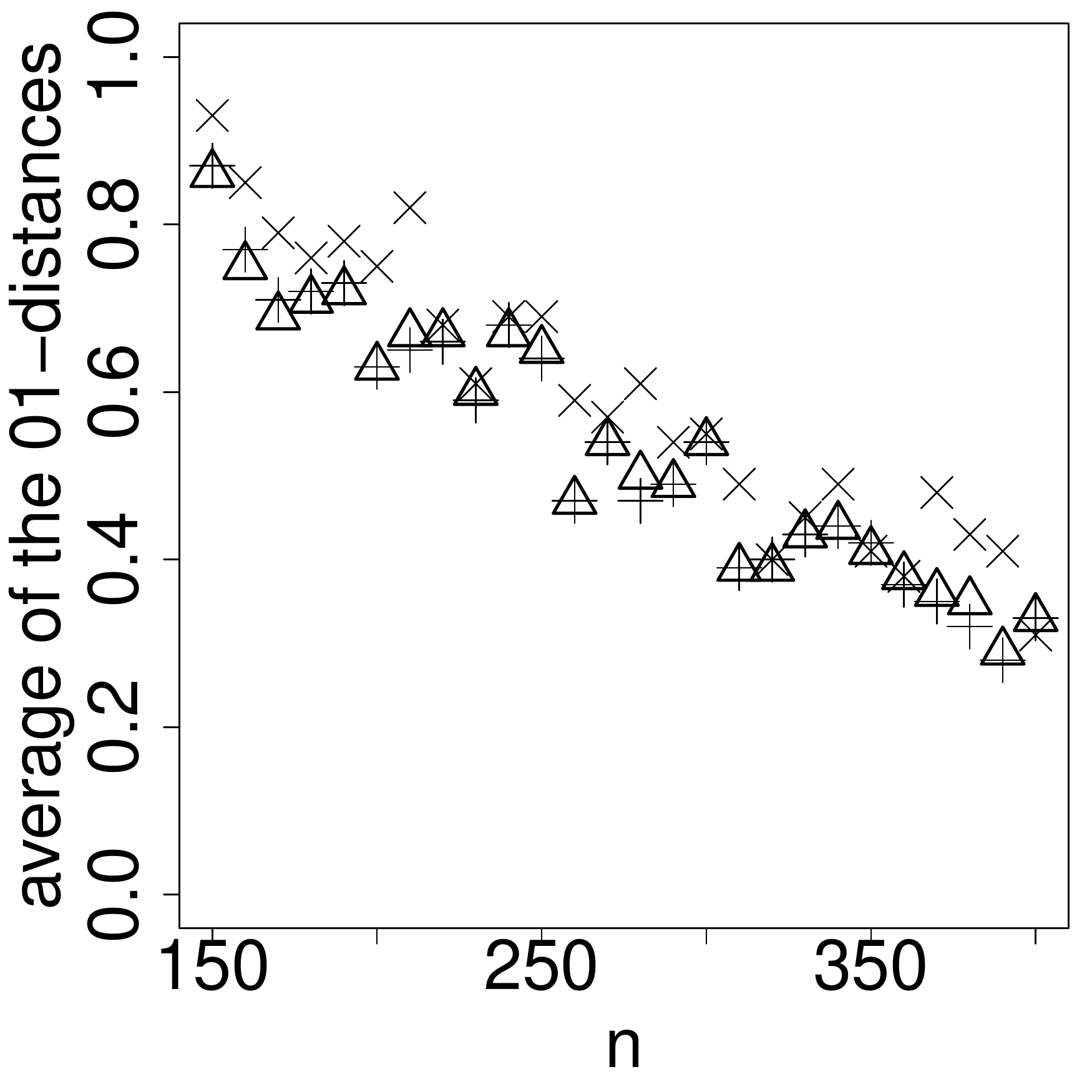}
&
\includegraphics[width=0.29\textwidth]{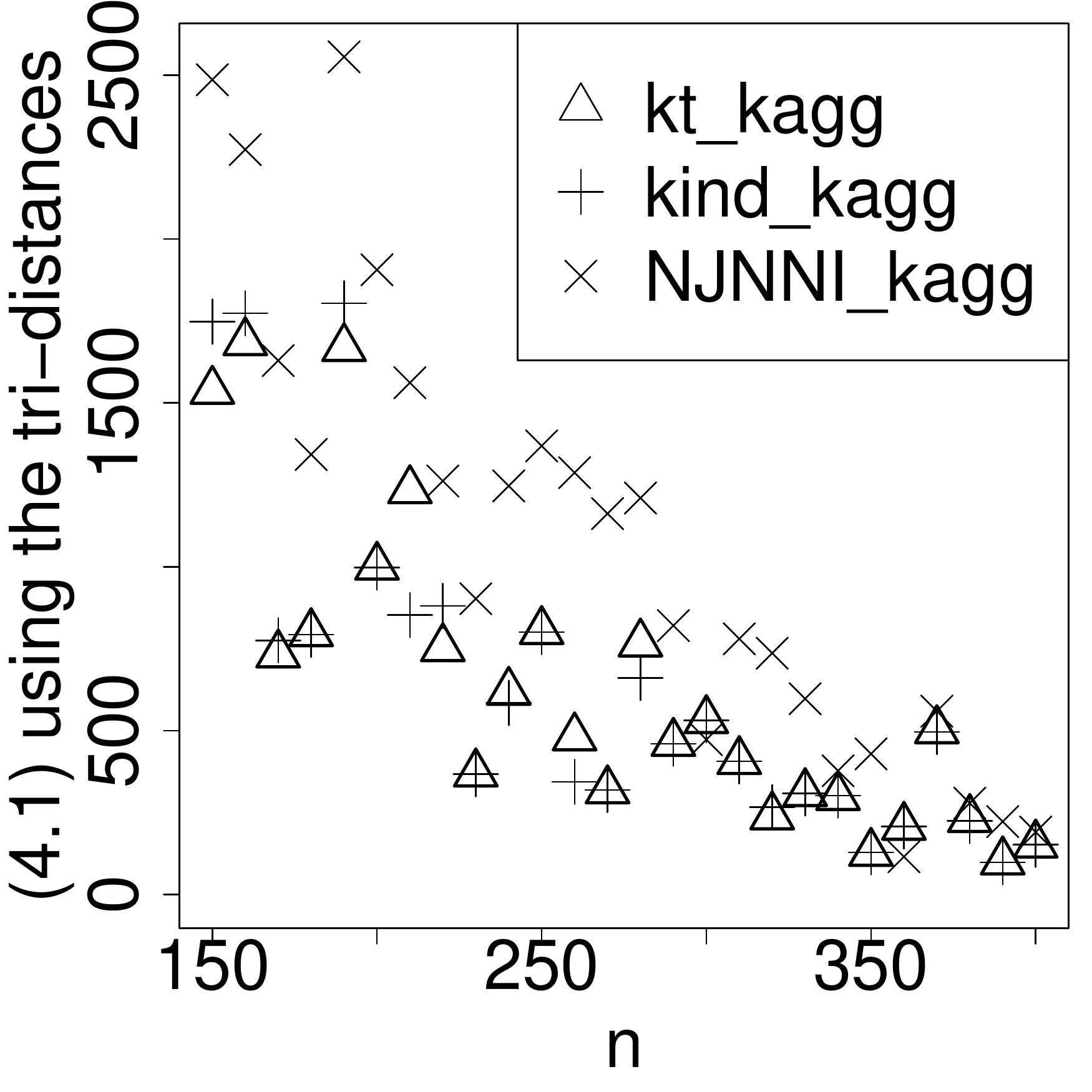}

\end{tabular}
\caption{results for the fifteenvariate structure. The generators used across the structure are all Joe generators. The set of parameters is $(\tau_{1:15}=0.1, \tau_{A_1}=0.25, \tau_{A_2}=0.5,\tau_{B_1}=0.5,\tau_{B_2}=0.75,\tau_{C_1}=0.35,\tau_{C_2}=0.45)$. \label{fift}}
\end{figure}

\begin{figure}[H]
\centering
\begin{tabular}{cc}
\includegraphics[width=0.29\textwidth]{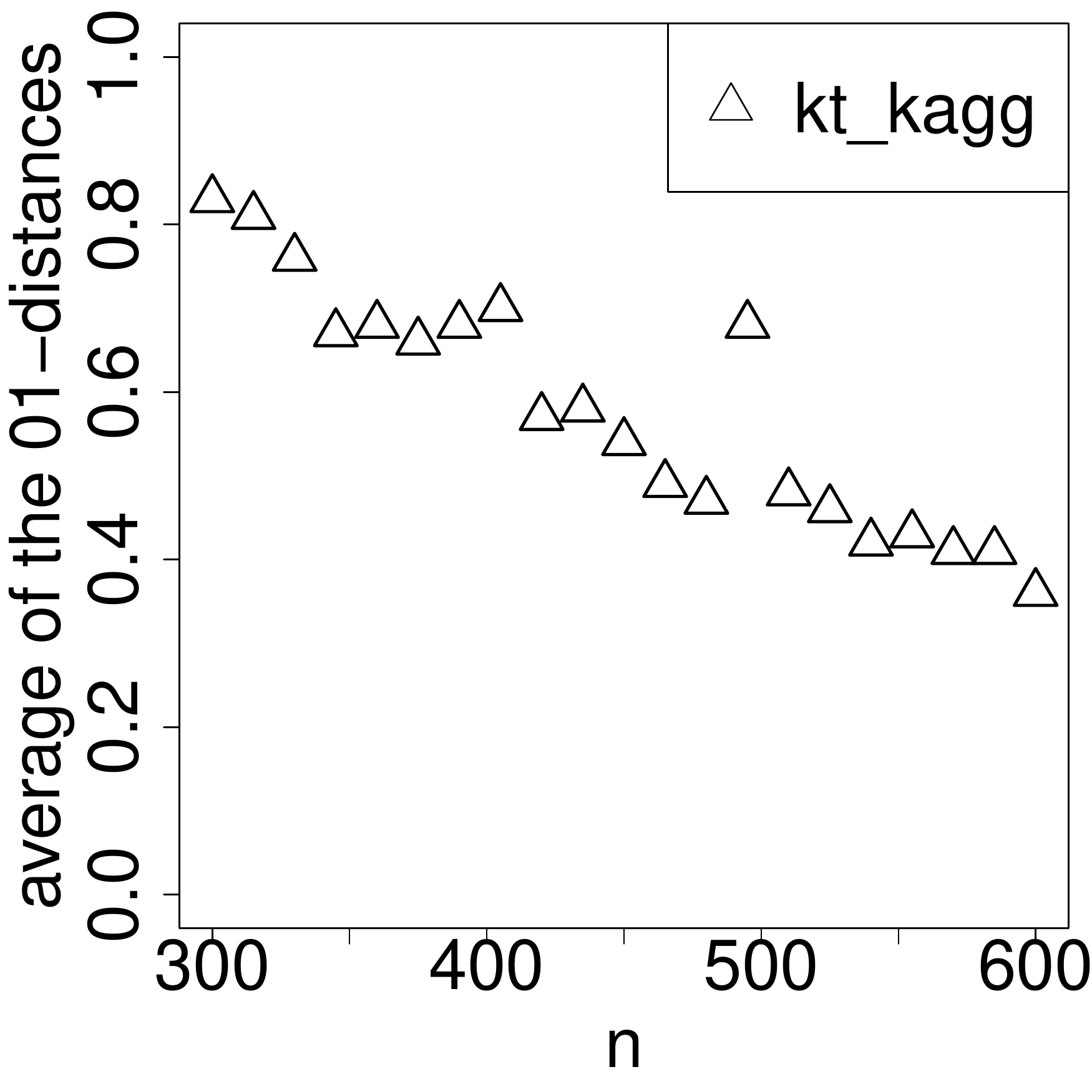}
&
\includegraphics[width=0.29\textwidth]{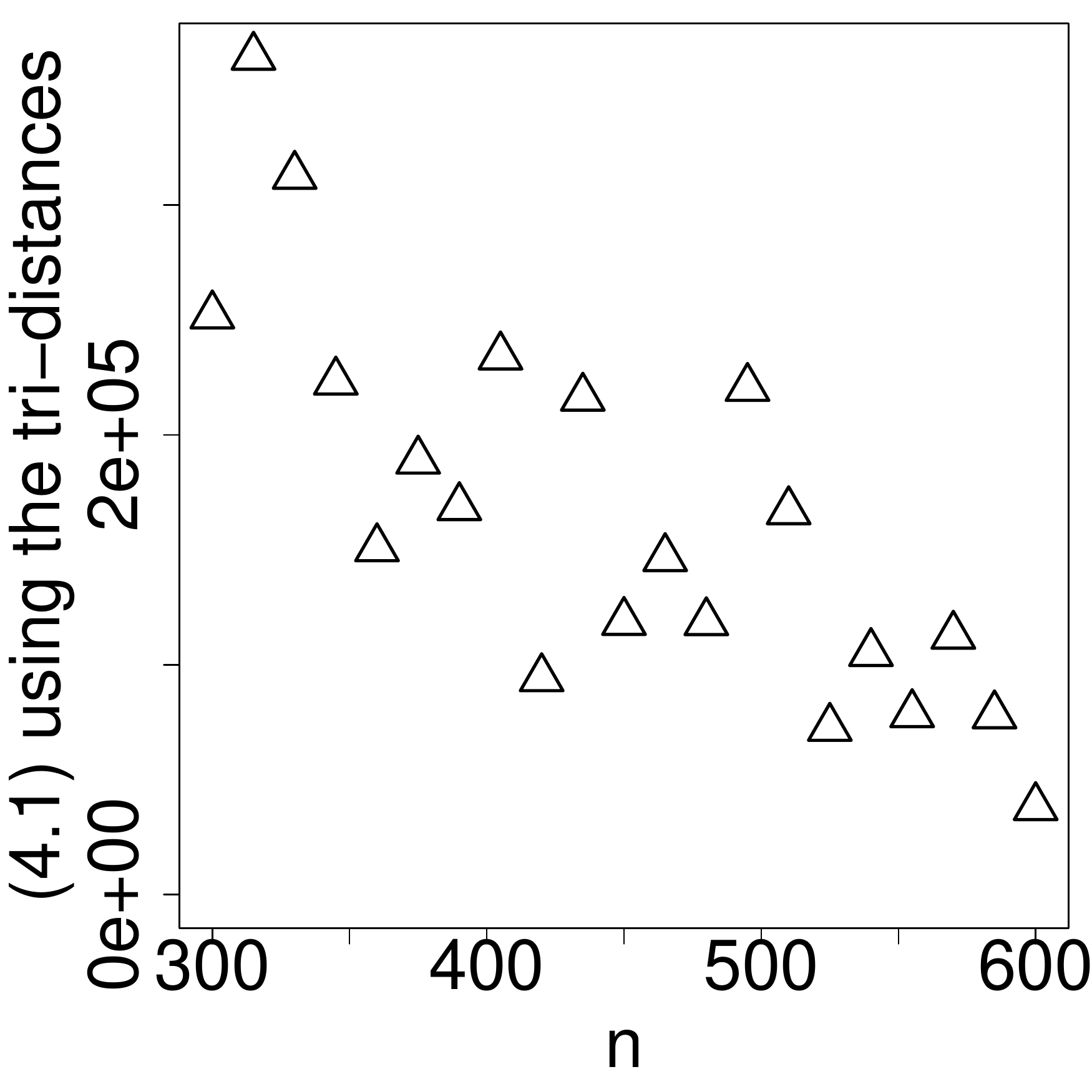}

\end{tabular}
\caption{results for the fortyvariate structure. The generators used across the structure are all Gumbel generators. The set of parameters is $(\tau_{1:40}=0.1,\tau_{A_1}=0.2,\tau_{A_2}=0.3,\tau_{A_3}=0.4,\tau_{A_4}=0.5,\tau_{A_5}=0.6,\tau_{A_6}=0.7,\tau_{A_7}=0.8,\tau_{B_1}=0.75,\tau_{C_1}=0.8,\tau_{D_1}=0.7,\tau_{E_1}=0.8,\tau_{F_1}=0.3,\tau_{F_2}=0.5,\tau_{F_3}=0.6,\tau_{G_1}=0.5,\tau_{G_2}=0.7,\tau_{H_1}=0.7)$. Only the \textsf{kt\_kagg} estimator was considered for this simulation, as this estimator turned out to be the fastest of them all. \label{perflarge}}
\end{figure}

Figure \ref{perf(12)(34)} through \ref{perflarge} give the simulation results for these target structures. The generators used across each structure and the related parameters, expressed as Kendall's $\tau$s for convenience, are specified below the figures.

Notice the average of the 01-distances is actually equal to the percentage of estimates that are unequal to the target structure, so that a value of 1 means not a single estimate of $\lambda$ among the $N=100$ available was equal to the target $\lambda$, while a value of 0 means all estimates of $\lambda$ were equal to $\lambda$.

Some comments are:
\begin{itemize}[noitemsep, nolistsep]
\item As the sample size increases, all estimators perform better.
\item The more resolved the target NAC is, the better the performance, as seen on Figure \ref{perf(12)(34)} through \ref{perfseven}. The tree structure of a poorly resolved NAC is, in general, harder to estimate.
\item Estimators from the new class are not necessarily a homogeneous group regarding performances (or execution times). For instance, at the top left of Figure \ref{perf(12)(34)}, one can see that the \textsf{NJNNI\_kb} estimator performs slightly better than the \textsf{S\&U} estimator in terms of mean 01-distance, while the \textsf{kt\_kagg} and \textsf{hD\_kagg} estimators both significantly perform better.
\item Several estimators from the new class beat the \textsf{S\&U} estimator by a large amount in case of binary target structures (Figure \ref{perf(12)(34)} and \ref{perfseven}). For instance, top right of Figure \ref{perfseven}, the \textsf{kt\_kagg} estimator can be seen to get the target structure wrong only 20\% of the time when the sample size is 30, while the \textsf{S\&U} estimator gets it wrong more than 80\% of the time on the same samples. Moreover, the bottom right part of the same figure shows that, when the \textsf{kt\_kagg} gets the target structure wrong, it's only by a very small, almost negligible, amount. When the \textsf{S\&U} estimator gets it wrong, the resulting estimate of $\lambda$ is usually far away from $\lambda$ itself.
\item In the simulations for the fourvariate binary structure, the \textsf{kt\_kagg} estimator turned out to be the one with the smallest execution times, producing estimates up to a 100 times faster than the \textsf{S\&U} estimator. Whatever the target NAC, estimators from the new class seemed to always exhibit smaller execution times than the \textsf{S\&U} estimator on the same data.
\end{itemize}

\section{Applications \label{application}}
In this section, a NAC tree structure is estimated on three different datasets, using the \textsf{kt\_kagg} estimator with $\tau_c$ set to 0.075. The working hypothesis is that each of these datasets is made of iid observations from a multivariate distribution with a nested Archimedean copula as copula. To help interpret a given estimated structure, an estimated summary of the generator at each internal node of the structure is obtained by averaging the estimated Kendall's $\tau$s of all the pairs of random variables interacting through that node.

The first dataset contains the results of 482 students to their exams. These students are in their first year, studying economics and management in a French-speaking Belgian university. They took 14 exams, that is, there are 14 random variables: Private Law, Psychology and Management, Sociology, Chemistry, Geography, English, Dutch, History, Mathematics, Physics, Statistics, a course designed to make sure students have the required prerequisites in sciences ($\text{BasicSci}$), and, finally, micro and macro economics (Econ103) and a related course (Econ104).

\begin{figure}[H]
\centering
\begin{minipage}[c]{1\textwidth}
\includegraphics[width=1\textwidth]{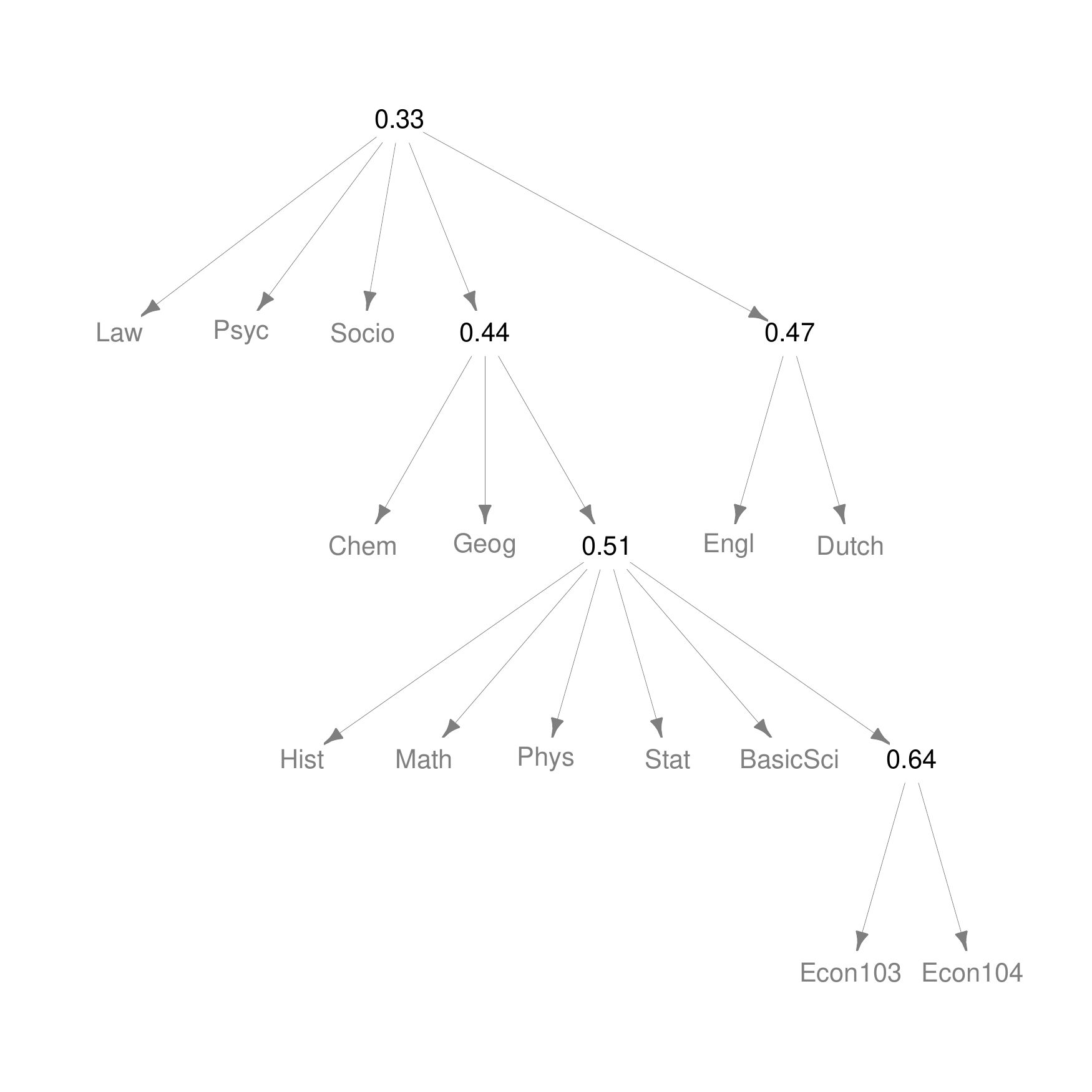}
\end{minipage}
\caption{NAC tree structure built on the grades of 482 students.\label{students}}
\end{figure}

Figure \ref{students} shows the estimated relationships between the various grades. The estimated mean Kendall's $\tau$ at the root is 0.33, suggesting that students tend to have good grades everywhere or bad grades everywhere, and less often a mix of good and bad grades. The strongest estimated Kendall's $\tau$ can be observed between the courses Econ103 and Econ104. Merging both courses in one exam could be a time-saving idea for the teachers in charge. English and Dutch courses are apart from the rest of the tree, with an estimated Kendall's $\tau$ of 0.47. Natural sciences such as Mathematics, Physics or Statistics are related through a rather strong estimated mean Kendall's $\tau$ (0.51), while courses such as Psychology or Sociology are both directly connected to the root where the dependence is the weakest.

The second dataset gives the percentage of workers employed in different sectors in European countries during 1979 (West and East). There are 26 countries and 9 random variables. The random variables are: percentage employed in Agriculture, in Mining, in Manufacturing, in Power Supply (PS), in Construction, in Service Industries (SI), in Finance, in Social and Personal Services (SPS), and, finally, in Transport and Communication (TC).

\begin{figure}[H]
\centering
\begin{minipage}[c]{0.425\textwidth}
\includegraphics[width=1\textwidth]{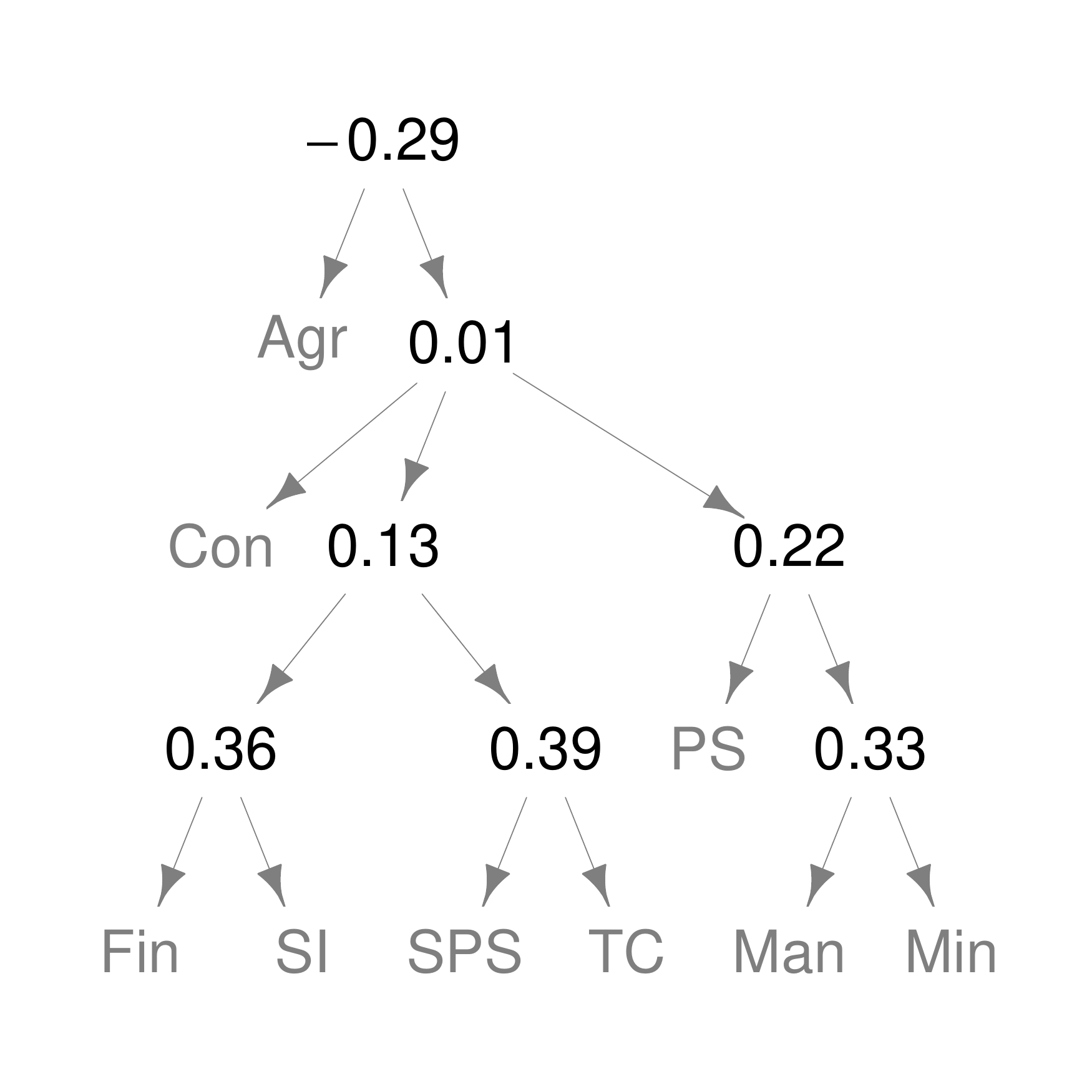}
\end{minipage}
\begin{minipage}[c]{0.55\textwidth}
\includegraphics[width=1\textwidth]{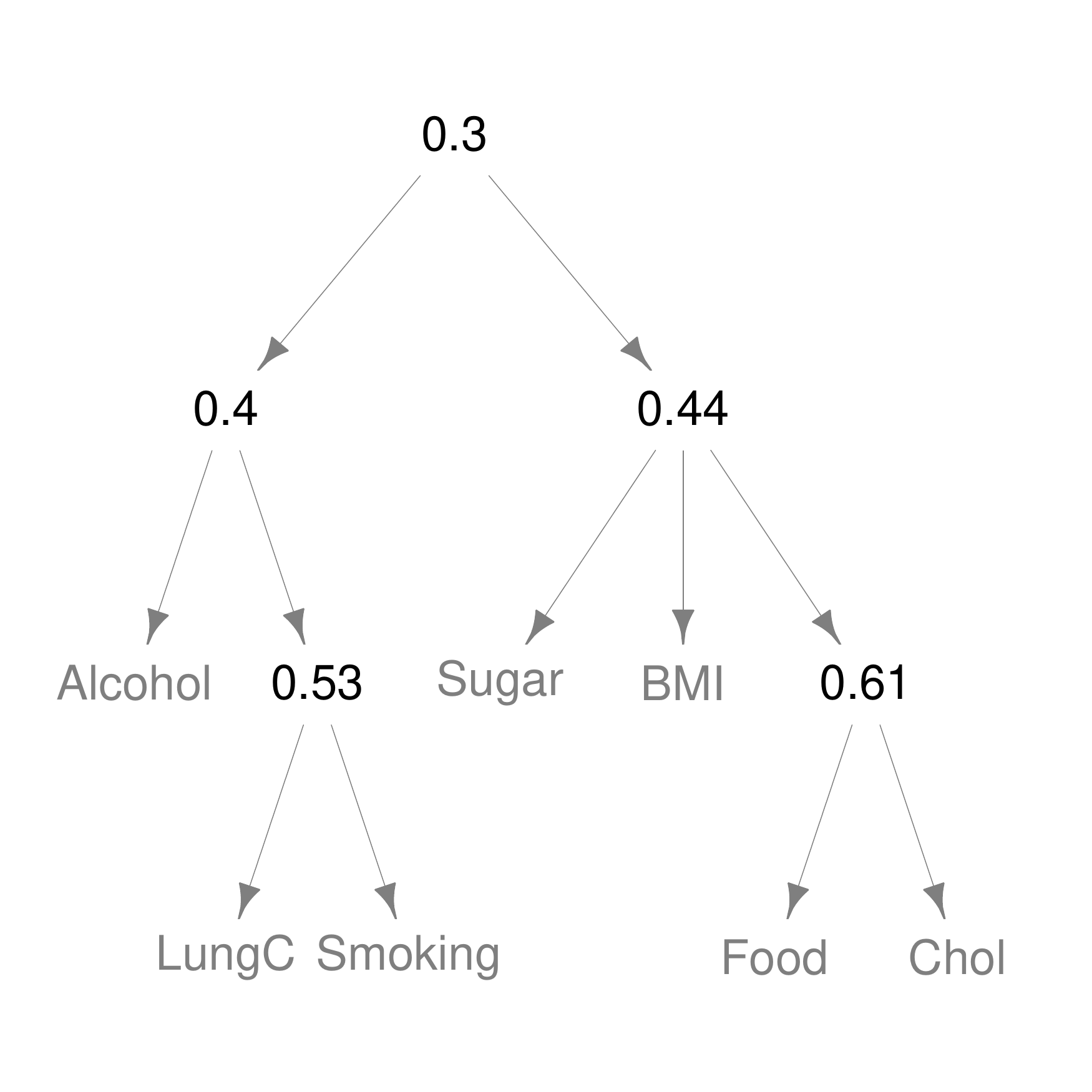}
\end{minipage}
\caption{Left: NAC tree structure built on 26 countries in West and East Europe, 1979. Right: NAC tree structure built on 104 countries across the world, 2002. \label{gapminder}}
\end{figure}

The estimated structure and its estimated mean Kendall's $\tau$s, displayed on the left-hand side of Figure \ref{gapminder}, offer insights about the relationships between sectors in terms of percentage of workers employed. A strong Manufacturing sector comes with a strong Mining sector and a lot of workers in the Power Supply sector as well. This bundle of sectors grows at the same time as the percentage of workers in the Agricultural sector decreases. Another bundle of sectors is made of the variables percentage of workers in the Finance sector, Service Industries, Social and Personal Services, and Transport and Communication. This branch of four variables, quite clearly the tertiary part of the economy, also grows at the same time as the percentage of workers in the Agricultural sector decreases.

The third dataset comes from the Gapminder Foundation\footnote{http://www.gapminder.org/data/}. The statistical units are 104 countries in 2002 from all over the world. The random variables are alcohol consumption by adult, new cases of lung cancer per 100,000 persons (LungC), prevalence of tobacco use by adult, sugar consumption per person, mean body mass index (BMI), number of calories available per person (Food) and mean total cholesterol (Chol).

The right-hand panel of Figure \ref{gapminder} shows a strong Kendall's $\tau$ between the prevalence of smoking and the number of new lung cancers observed, which was expected. Alcohol is not far. The number of calories and the amount of sugar available are related to the body mass index but also to the cholesterol in the blood, no surprise there either.

In all three cases, only the tree structure of the target NAC has been estimated. However, if one would make the rather strong assumption that the generators across a given estimated structure are all from the Clayton family (or another known family of generators), the only thing left to estimate would be a set of parameters, one for each generator. This can be done using the maximum likelihood approach or, in case we assume Clayton generators across the structure, by making use of the estimated mean Kendall's $\tau$s and of the equation
\begin{equation} \nonumber
\theta = 2/(\tau^{-1}-1),
\end{equation}
where $\theta$ is the parameter related to a Clayton generator (\citealp{Hofert:Maechler:2010:JSSOBK:v39i09}).

Unfortunately, dealing with the generators this way does not ensure the resulting estimated NAC will always be a proper copula. This is discussed in more details in the next section.

\section{Discussion: the future of NACs \label{discussion}}
As mentioned in the introductory section, ACs have become a standard tool for modelling or simulating bivariate data, but their success story falls short as soon as they are used to model or simulate higher-dimensional datasets. Since NACs offer more flexibility for modelling dependencies in a higher-dimensional setting while still reducing to Archimedean copulas in simpler cases, it is strongly believed by the author of this paper that the class of nested Archimedean copulas has the potential to replace the class of Archimedean copulas in the future as an even more popular tool for simulation and modelling purposes.

Nested Archimedean copulas are unfortunately not yet ready for this to happen. While it is known that the tree structure of a NAC must be a rooted phylogenetic tree, a sufficient and necessary condition on the generators to ensure the resulting NAC to be a proper copula is, at the time of writing, still unknown. As long as this sufficient and necessary condition will remain unknown, the estimation of the generators is very likely to remain an issue as well. This is reflected in the only two existing papers dealing with the estimation of NAC tree structures:

\begin{itemize}[noitemsep, nolistsep]
\item \cite{OOW} offer what could be described as a parametric approach for estimating NACs. In their approach, each generator is assumed to be known up to one or several Euclidean parameters and all generators are assumed to belong to the same parametric family. The set of parameters and the tree are then estimated in such a way that the resulting NAC will fulfill the sufficient (but not necessary) nesting condition developed by \cite*{tJOE97a} and \cite*{doi:10.1080/00949650701255834}, this condition ensuring the estimated NAC to be a proper copula. The choice of a common parametric family for the generators remains a serious problem.
\item \cite{segers2014nonparametric} developed a method to estimate the tree structure of a NAC. In contrast to the method from \cite{OOW}, nothing is assumed about the target NAC prior to the estimation of its tree. The problem of estimating the generators is left to the reader.
\end{itemize}

Estimation of the generators of a target nested Archimedean copula one generator at the time and independently from the other generators does not ensure the resulting estimated NAC to be a proper copula, as estimation of each generator independently from the other generators does not guarantee to result in a set of generators meeting the unknown sufficient and necessary condition that must apply on these generators. Estimation of each generator (or of a summary measure of each generator) independently from the other generators can however be very useful for deciding if any two successive nodes of an estimated tree should be collapsed into one to avoid overfitting (Section~\ref{new_est}) or for interpretation of an estimated tree structure (Section~\ref{application}).


In conclusion, researchers sharing the feeling nested Archimedean copulas could make it as a popular class of copulas are strongly invited to work on the remaining issues to make this happen. What is a sufficient and necessary condition on the generators of a NAC? How can these generators be estimated and how to make sure they will fulfill this sufficient and necessary condition? Once a NAC has been fully estimated (tree + nonparametric generators), how to generate new observations from that estimated NAC? What about goodness-of-fit tools? These issues must be addressed.


\section*{Acknowledgements}
The author of this paper would like to deeply thank Johan Segers (Universit\'e catholique de Louvain) for his many constructive comments and suggestions regarding this research.

Special thanks also go to Liam J. Revell (University of Massachusetts) for always replying to my mails with detailed answers and for his general patience in explaining phylogenetics to a statistician.

Finally, the author of this paper is grateful for support of this research by both the ``Communaut\'e fran\c{c}aise de Belgique'' through contract ``Projet d'Actions de Recherche Concert\'ees'' No.\ 12/17-045 and the Belgian government (Belgian Science Policy) through IAP research network grant nr. P7/06.

\bibliographystyle{plainnat}
\bibliography{uyttendaele_nested_archimedean_copulas}

\end{document}